\documentclass[twocolumn]{aastex61}
\shorttitle{Stellar Parameter and Abundance Comparisons With Independent Analyses}
\shortauthors{J\"onsson et al.}

\usepackage{textcomp}

\begin{document}

\title{APOGEE Data Releases 13 and 14:\ \\Stellar Parameter and Abundance Comparisons With Independent Analyses}

\author[0000-0002-4912-8609]{Henrik J\"onsson}
\affil{Instituto de Astrof\'isica de Canarias (IAC), E-38205 La Laguna, Tenerife, Spain}
\affil{Universidad de La Laguna, Dpto. Astrof\'isica, E-38206 La Laguna, Tenerife, Spain}
\affil{Lund Observatory, Department of Astronomy and Theoretical Physics, Lund University, Box 43, SE-22100 Lund, Sweden}

\author{Carlos Allende Prieto}
\affil{Instituto de Astrof\'isica de Canarias (IAC), E-38205 La Laguna, Tenerife, Spain}
\affil{Universidad de La Laguna, Dpto. Astrof\'isica, E-38206 La Laguna, Tenerife, Spain}

\author[0000-0002-9771-9622]{Jon A. Holtzman}
\affil{New Mexico State University, Las Cruces, NM 88003, USA}

\author[0000-0002-3101-5921]{Diane K. Feuillet}
\affil{Max Planck Institute for Astronomy, K\"onigstuhl 17, 69117 Heidelberg, Germany}

\author[0000-0002-1423-2174]{Keith Hawkins}
\affil{Columbia University, 550 West 120th Street, New York 10027, USA}
\affil{Department of Astronomy, the University of Texas at Austin, 2515 Speedway Boulevard, Austin, TX 78712, USA}

\author{Katia Cunha}
\affil{Steward Observatory, The University of Arizona, 933 North Cherry Avenue, Tucson, AZ 85721-0065, USA}
\affil{Observat\'orio Nacional, Rua General Jos\'e Cristino, 77, 20921-400 S\~ao Crist\'ov\~ao, Rio de Janeiro, RJ, Brazil}

\author{Szabolcs M\'esz\'aros}
\affil{ELTE E\"otv\"os Lor\'and University, Gothard Astrophysical Observatory, Szombathely, Hungary}
\affil{Premium Postdoctoral Fellow of the Hungarian Academy of Sciences}

\author{Sten Hasselquist}
\affil{New Mexico State University, Las Cruces, NM 88003, USA}

\author{J. G. Fern\'andez-Trincado}
\affil{Departamento de Astronom\'\i a, Casilla 160-C, Universidad de Concepci\'on, Concepci\'on, Chile}
\affil{Institut Utinam, CNRS UMR6213, Univ. Bourgogne Franche-Comt\'e, OSU THETA, Observatoire de Besan\c{c}on, BP 1615, 25010 Besan\c{c}on Cedex, France}

\author{D. A. Garc\'ia-Hern\'andez}
\affil{Instituto de Astrof\'isica de Canarias (IAC), E-38205 La Laguna, Tenerife, Spain}
\affil{Universidad de La Laguna, Dpto. Astrof\'isica, E-38206 La Laguna, Tenerife, Spain}

\author{Dmitry Bizyaev}
\affil{Apache Point Observatory and New Mexico State University, PO Box 59, Sunspot, NM 88349-0059, USA}
\affil{Sternberg Astronomical Institute, Moscow State University, Moscow, Russia}

\author[0000-0001-6143-8151]{Ricardo Carrera}
\affil{INAF-Osservatorio astronomico di Padova, Vicolo dell'Osservatorio 5, I-35122 Padova, Italy}

\author{Steven R. Majewski}
\affil{Department of Astronomy, University of Virginia, Charlottesville, VA 22904-4325, USA}

\author[0000-0002-7549-7766]{Marc H. Pinsonneault}
\affil{Department of Astronomy, The Ohio State University, 140 West 18th Avenue, Columbus OH 43210, USA}

\author{Matthew Shetrone}
\affil{University of Texas at Austin, McDonald Observatory, Fort Davis, TX 79734, USA}

\author{Verne Smith}
\affil{National Optical Astronomy Observatory, 950 North Cherry Avenue, Tucson, AZ 85719, USA}

\author{Jennifer Sobeck}
\affil{Department of Astronomy, Box 351580, University of Washington, Seattle, WA 98195, USA}

\author[0000-0002-7883-5425]{Diogo Souto}
\affil{Observat\'orio Nacional, Rua General Jos\'e Cristino, 77, 20921-400 S\~ao Crist\'ov\~ao, Rio de Janeiro, RJ, Brazil}

\author[0000-0003-1479-3059]{Guy S. Stringfellow}
\affil{Center for Astrophysics and Space Astronomy, Department of Astrophysical and Planetary Sciences, University of Colorado, Boulder, CO, 80309-0389, USA}

\author{Johanna Teske}
\affil{Hubble Fellow, Carnegie Department of Terrestrial Magnetism, 5241 Broad Branch Road, NW, Washington, DC 20015-1305}
\affil{Carnegie Origins Fellow, jointly appointed by Carnegie DTM/OBS}

\author{Olga Zamora}
\affil{Instituto de Astrof\'isica de Canarias (IAC), E-38205 La Laguna, Tenerife, Spain}
\affil{Universidad de La Laguna, Dpto. Astrof\'isica, E-38206 La Laguna, Tenerife, Spain}



\begin{abstract}
Data from the SDSS-IV / Apache Point Observatory Galactic Evolution Experiment (APOGEE-2) have been released as part of SDSS Data Releases 13 (DR13) and 14 (DR14). These include high resolution $H$-band spectra, radial velocities, and derived stellar parameters and abundances. DR13, released in August 2016, contained APOGEE data for roughly 150,000 stars, and DR14, released in August 2017, added about 110,000 more. Stellar parameters and abundances have been derived with an automated pipeline, the APOGEE Stellar Parameter and Chemical Abundance Pipeline (ASPCAP). We evaluate the performance of this pipeline by comparing the derived stellar parameters and abundances to those inferred from optical spectra and analysis for several hundred stars.  For most elements -- C, Na, Mg, Al, Si, S, Ca, Cr, Mn, Ni -- the DR14 ASPCAP analysis have systematic differences with the comparisons samples of less than 0.05 dex (median), and random differences of less than 0.15 dex (standard deviation). These differences are a combination of the uncertainties in both the comparison samples as well as the ASPCAP-analysis. Compared to the references, magnesium is the most accurate alpha-element derived by ASPCAP, and shows a very clear thin/thick disk separation, while nickel is the most accurate iron-peak element (besides iron).
\end{abstract}




\section{Introduction}\label{sec:introduction}
The Apache Point Observatory Galactic Evolution Experiment (APOGEE-2) is an ongoing project within SDSS-IV \citep{2017AJ....154...94M,2017AJ....154...28B,2011AJ....142...72E,2017ApJS..233...25A,2018ApJS..235...42A,2013AJ....146...81Z,2017AJ....154..198Z,2015AJ....150..173N,2006AJ....131.2332G,2012SPIE.8446E..0HW}, analyzing spectroscopically stars of all major galactic components using H-band spectra (R$\sim$22,500). The reduced spectra, information about the observations, and the stellar parameters and stellar abundances determined from the spectra are periodically released to the public. In \citet{paperi}, the two most recent data releases -- DR13 from July 2016 and DR14 from August 2017 -- are presented. DR13 contains 164,562 stars  observed between April 2011 and July 2014, while DR14 contains 277,371 stars observed between April 2011 and January 2016. Currently, the APOGEE analysis include 22 elements, meaning that the APOGEE dataset is a unique dataset for astronomical research based on tracing chemical evolution of stellar populations and/or chemical tagging of stars. In this paper we attempt to assess the accuracy of the DR13 and DR14 APOGEE stellar parameters and abundances through a comparison to alternate analyses of a subset of APOGEE stars. 

The accuracy of the APOGEE stellar parameters and abundances have been examined several times before:

\citet{2016A&A...594A..43H} made an independent spectroscopic analysis of the APOGEE spectra to determine abundances of C, N, O, Na, Mg, Al, Si, S, K, Ca, Ti, V, Cr, Mn, Co, and Ni in the 2012 giant stars from DR12 that also have asteroseismic analysis from Kepler light curves, which provides constraints on the stellar parameters. Their determined abundances show small differences to that of DR12 for many elements, but significant differences regarding Si, S, Ti, and V.

\citet{2016ApJ...830...35S} use the DR13 APOGEE spectra to determine C, N, O, Na, Mg, Al, Si, K, Ca, Ti, V, Cr, Mn, Co and Ni in twelve giant stars within the open cluster NGC 2420 ([Fe/H]$\sim -0.16$). The results from this manual analysis compare well with DR13 abundances for most of the elements studied, although for Na, Al, and V there are larger differences.

\citet{2016ApJ...833..132F} manually re-analyzed the DR12 APOGEE spectrum of one peculiar metal-poor field giant star with a globular cluster (GC) second-generation (SG) abundance pattern. They derived the stellar parameters using iSpec \citep{2014A&A...569A.111B} and found values very close to the APOGEE DR12 values. For the abundances C, N, O, Mg, and Al, they found differences of about 0.3 dex when comparing their manually, MOOG\footnote{\url{http://www.as.utexas.edu/~chris/moog.html}}-derived abundances to those of DR12. Subsequently, the same star was analyzed using an optical spectrum by \citet{2017MNRAS.469..774P}, confirming its SG abundance pattern, but their derived stellar parameters differ significantly from those of the APOGEE analysis pipeline: they arrive at an effective temperature that is 300 K lower than DR14, and a surface gravity that is 0.9 dex lower\footnote{In fact, the APOGEE analysis pipeline is concluded to be not very precise for SG stars with extreme `non-standard' abundance patterns in Section \ref{sec:sg}.}. The metallicity is however very similar in \citet{2017MNRAS.469..774P} and DR14. More DR13 APOGEE spectra of SG-type field stars have been analyzed manually in \citet{2017ApJ...846L...2F}, and the derived abundances generally agree within 0.2 dex with the abundances in DR13.  

\citet{2017ApJ...835..239S} used the DR13 APOGEE spectra to determine C, O, Na, Mg, Al, Si, K, Ca, Ti, V, Cr, and Mn in two planet-hosting M-dwarf stars. This work shows that APOGEE spectra can be analyzed to determine detailed chemical compositions of M-dwarfs, if FeH is included in the analysis. Since this molecule is not included in the DR13 or DR14 line lists, this work concludes that no results from these data releases regarding M-dwarfs can be fully trusted, but that this issue may be solved in upcoming data releases; the plan is to use FeH in future APOGEE analysis.

\citet{2018AJ....155...68W} compare the effective temperatures and metallicities as derived by the APOGEE stellar parameter and chemical abundances pipeline (ASPCAP, DR14) for 221 dwarf stars from the Kepler Object of Interest catalogue to those derived from independent, optical analyses and find the DR14 effective temperatures to be around 60 K lower than the optically determined effective temperatures in all their reference samples, with a spread of around 130 K. Regarding [Fe/H], they find a zero mean offset and a spread of 0.09 dex.

There have also been several other works in which one or more stellar parameter(s) have been independently determined and either have been, or could be, compared to those derived by APOGEE; for example, the APOKASC-project \citep{2014ApJS..215...19P} has derived asteroseismic surface gravities for 1916 red giants based on data from Kepler and APOGEE \citep[as used by][]{2016A&A...594A..43H}. This catalog have recently been updated to include 6,681 targets \citep{Pinsonneault:2018vh}. However, these surface gravities are subsequently used to calibrate the surface gravities in DR13 and DR14, and can therefore not be considered as an independent analysis. The same type of asteroseismic-spectroscopic analysis has been made on 606 stars observed by CoRoT and APOGEE within the CoRoGEE-project \citep{2017A&A...597A..30A}. Also short-time variations -- `flicker' -- in the light-curves of Kepler targets have been used as a basis for determining surface gravities \citep{2016ApJ...818...43B}, and there are of course several photometric calibration relations designed to estimate effective temperatures of stars \citep[e.g.,][]{2009A&A...497..497G}.

Furthermore, the data-driven analysis code called the Cannon \citep{2015ApJ...808...16N,2016arXiv160303040C} has been used on APOGEE spectra to determine stellar parameters and stellar abundances for a majority of the DR14 APOGEE-sample of stars (see \citet{paperi} for details). However, since this analysis is based on a training-set from the ASPCAP DR14 results, this analysis cannot be considered independent of the DR14-values. In fact, these results will be evaluated in this paper in addition to the ASPCAP derived stellar parameters and abundances.

None of the works above have compared a large number of stars of any APOGEE release to reference analyses independently determining all the classical spectroscopic stellar parameters -- effective temperature, surface gravity, and metallicity -- as well as at least some elemental abundances. This paper conducts a deeper analysis of the accuracy of the APOGEE DR13/DR14 stellar parameters and abundances by comparing these results to those of sizable independent studies.

\section{The APOGEE DR13 and DR14 samples}\label{sec:apogee}

The spectral analysis that determines the stellar parameters and chemical abundances is performed automatically by ASPCAP \citep{2016AJ....151..144G}. The stellar parameters of a particular star are determined by optimization using a large library of pre-computed synthetic spectra with different stellar parameters, C, N and overall alpha elemental abundances covering the entire APOGEE wavelength range 15,140-16,940~\AA~\citep{2015AJ....149..181Z}. The same spectral library is then used with the determined stellar parameters fixed (to the uncalibrated values, see Section \ref{sec:params}) to derive the abundances of the individual elements using windows corresponding to spectral lines that are sensitive to the element of interest. The determination of stellar parameters and abundances is made with the code FERRE\footnote{Available at \url{http://github.com/callendeprieto/ferre}} \citep{2006ApJ...636..804A} and the model atmospheres used are MARCS \citep{2008A&A...486..951G} plane-parallel/spherical models (for high/low $\log g$) for $T_{\mathrm{eff}} < 3500$ K and ATLAS9 \citep[][and updates]{1979ApJS...40....1K} plane-parallel models for $T_{\mathrm{eff}} \ge 3500$ K \citep{2012AJ....144..120M}. The spectral libraries in DR13 and DR14 were calculated with Turbospectrum \citep{1998A&A...330.1109A,2012ascl.soft05004P} using plane-parallel or spherical radiative transfer consistently with the stellar model in question. The same line list is used to construct the synthetic spectral libraries for both DR13 and DR14 (internally tagged as 20150714). It is based on a thorough literature review in combination with astrophysical $\log gf$-values determined using high resolution atlas-spectra of the Sun and Arcturus \citep{1991aass.book.....L,1995iaas.book.....H}; for more details, see \citet{2015ApJS..221...24S}. The elements analyzed and released in DR13 and DR14 are the same: C, N, O, Na, Mg, Al, Si, P, S, K, Ca, Ti, V, Cr, Mn, Co, Ni, Cu, Ge, Rb, Nd, and Yb. However, no comparisons for Cu, Ge, Rb, Nd, and Yb are made in this paper, mainly since the determinations of these elements is a work in progress, and still not fully reliable, see Section \ref{sec:nc}.

For some of the observed stars, ASPCAP fails to determine the stellar parameters for one reason or another (the S/N could be very low, the star could be too cool/hot, the star could be a spectroscopic binary, etc.), which means that 152,641 stars (93\%) have stellar parameters in DR13 and 264,078 stars (95\%) have stellar parameters in DR14. 

The S/N distributions in the two data releases are similar, with DR14 having a slightly higher fraction of high-S/N spectra: in DR13 29\% have S/N$<100$ and for DR14 the fraction is 26\%. The flag SN\_WARN is triggered for stars with S/N$<70$ and the flag SN\_BAD is triggered for stars with S/N$<30$.

The `raw' output from FERRE is calibrated to reproduce surface gravities determined by asteroseismology, to yield homogeneous abundances in clusters of stars, and to reproduce solar abundance ratios for stars with near-solar metallicity in the solar vicinity (for details, see \citet{paperi}). Because of the lack of asteroseismic surface gravities for dwarfs at the time of data release calibration for DR13 and DR14, only stars with $\log g<4.0$ have calibrated surface gravities. As a consequence the giant and subgiant stars with $\log g<4.0$ (105,599 in DR13 and 159,047 in DR14) have more accurate ASPCAP-parameters than the dwarf stars with $\log g>4.0$.
 
\begin{figure*}
\epsscale{1.18}
\plotone{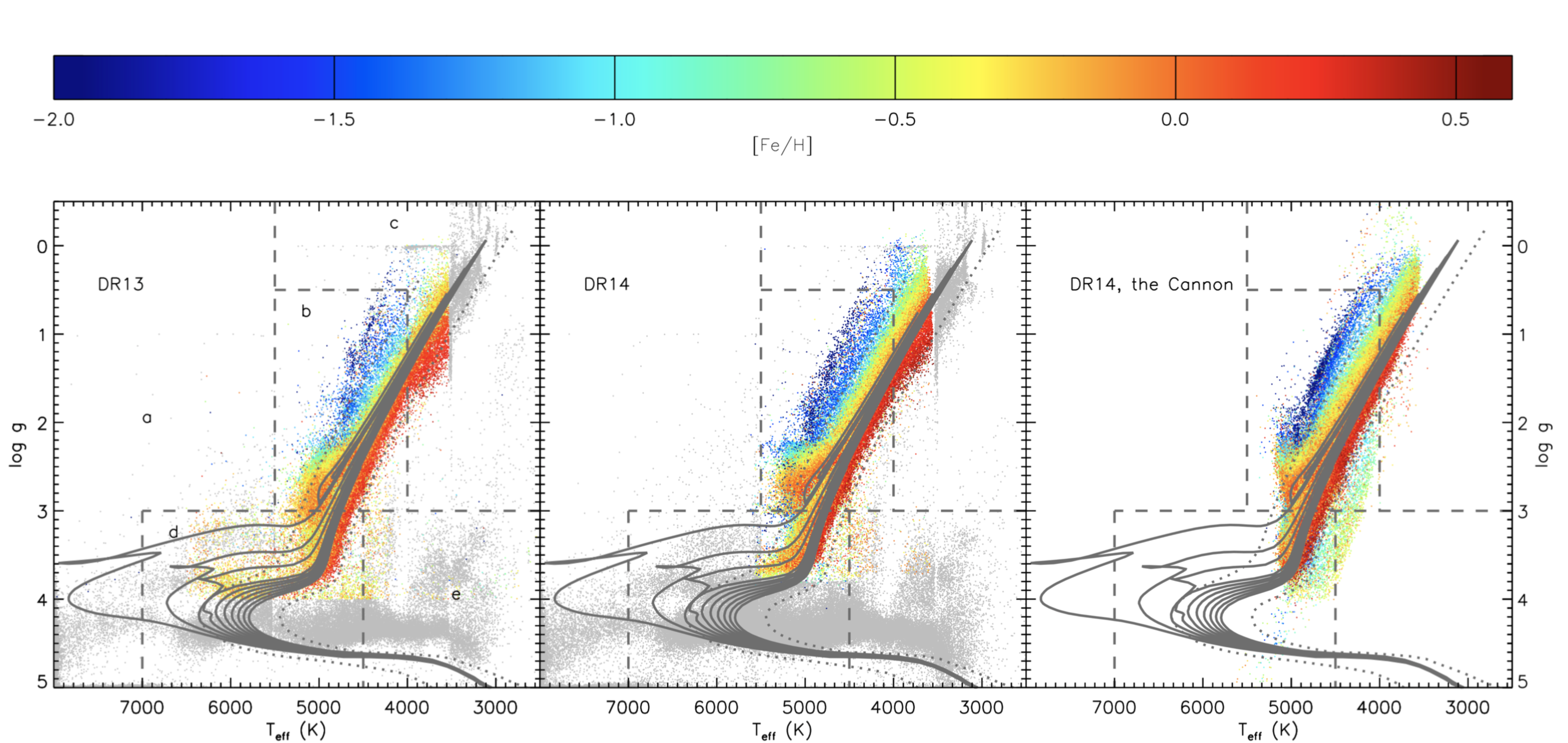}
\caption{HR diagrams for the ASPCAP-analyses of DR13 and DR14, as well as the DR14-based Cannon-analysis. Stars that have calibrated values for all three `classical' spectroscopic stellar parameters are plotted color-coded according to their metallicity, while stars with one or more uncalibrated stellar parameter are plotted in gray. Crude regions of different types of stars are marked: hot stars (a), GK-giants (b), M-giants (c), FGK-dwarfs (d), and KM-dwarfs (e). As a guide for the eye, isochrones with [Fe/H]=0.0 and ages 1-10 Gyr are plotted using solid dark gray lines. Furthermore, one isochrone with [Fe/H]=-1.0 and age 10 Gyr, and one with [Fe/H]=+0.5 and age 10 Gyr are plotted using dotted dark gray lines \citep{2012MNRAS.427..127B}.\label{fig:hr-apogee}}
\end{figure*}

The HR-diagrams based on the DR13 and DR14 ASPCAP-analyses as well as the DR14-based Cannon-analysis are shown in Figure \ref{fig:hr-apogee}. The stars with calibrated values for all three `classical' spectroscopic stellar parameters (effective temperature, surface gravity, and metallicity) are color-coded with respect to their iron-abundances, and the giant branch lines up just as expected from isochrones in effective temperature, surface gravity, and metallicity. Stars with one or more uncalibrated stellar parameter(s) are shown in gray. Also shown is a very crude division into different types of stars based on `typical', optical spectroscopic reference samples as described in Section \ref{sec:indep}. From this division, the DR13-sample consists of 6\% hot stars, 46\% GK-giants, 12\% M-giants, 26\% FGK-dwarfs, and 10\% KM-dwarfs (using uncalibrated T$_{\mathrm{eff}}$ and $\log g$); the same values for DR14 are 6\%, 42\%, 9\%, 34\%, and 10\%, respectively. Hence, DR14 has a larger fraction of FGK-dwarfs and a lower fraction of GK-giants as compared to DR13.

To make the comparison between ASPCAP and other independent analyses in this paper as relevant as possible, we have chosen to compare only stars that are not flagged to have an uncertain or bad ASPCAP analysis \citep[see][for a description of the APOGEE flags]{2015AJ....150..148H}, and which have calibrated values for all three stellar parameters in DR13/DR14. For example, we know from \citet{2017ApJ...835..239S} that the DR13 and DR14 versions of ASPCAP are not producing reliable results for M-dwarfs due to a lack of FeH molecular lines in the adopted synthetic spectra, and from Figure \ref{fig:hr-apogee}, we see that the results for dwarfs from ASPCAP are not following the main sequence expected from isochrones. This means that we are left with the subgiant and GKM-giant stars, but this is still a majority of the APOGEE sample: 105,599 stars of the 152,641 stars with determined parameters in DR13 have calibrated values for all three stellar parameters (69\%), while 159,047 of the 264,078 stars with determined parameters in DR14 have calibrated values for all three stellar parameters (60\%).

\section{Independent analyses}\label{sec:indep}
We have surveyed the literature for suitable independent works with which to compare the ASPCAP determined stellar parameters and abundances. In this section we describe the reference samples. We focus mainly on works that have a significant number of stars ($\gtrsim100$), which have all classical stellar parameters -- effective temperature, surface gravity, and metallicity -- determined together with as many elemental abundances as possible, but we also discuss some other smaller works of special interest.

\subsection{Field star samples}\label{sec:ref}
When comparing stellar parameters and abundances between different works it is hard to say which one is most accurate, since ``ground truth'' usually is not available. Therefore, comparing two samples where the overlap is just a few stars is not expected to say much of interest on a statistical basis: it is desirable to be able to distinguish possible systematic trends of differences as functions of, for example, metallicity, effective temperature, etc. from differences stemming from the combined random uncertainties.
Due to the small number of overlapping stars with APOGEE, we did not use the samples of stars analyzed by \citet{2008A&A...487..373S,2011A&A...533A.141S,2011A&A...526A..99S,2012A&A...545A..32A,2013A&A...555A.150T,2016A&A...591A..69S,2017A&A...606A..94D} and \citet{2014A&A...562A..71B,2015A&A...577A...9B,2016A&A...586A..49B} and \citet{1998A&A...338..161F,Anonymous:abKu9BK8,2004AN....325....3F,2008MNRAS.384..173F,2011MNRAS.414.2893F,2012ApJS..203...30F,2015ApJ...809..107F,2017MNRAS.464.2610F} and \citet{2017MNRAS.468.4151I} and \citet{2003MNRAS.340..304R,2006MNRAS.367.1329R} and \citet{2005ApJS..159..141V} in the comparison. The small numbers of targets overlapping between these studies and APOGEE is not surprising, since these works all are mainly based on dwarf stars, and APOGEE is mainly targeting giant stars\footnote{Evaluating the ASPCAP performance in the uncalibrated regime of the FGK-dwarfs, by comparing to reference samples containing a high degree of such stars will be performed in the future (Teske et al., in prep).}. Furthermore, we considered comparing to the GALAH survey, but it was not used because the overlap with their first data release included only 23 stars \citep{2017MNRAS.465.3203M}. A very interesting study for comparison with APOGEE is the metal-poor giant star sample of \citet{2011ApJ...737....9R}, but unfortunately only one star from this sample is presently among the APOGEE-observed stars. Interesting samples, in spite of their small numbers, are the Gaia benchmark stars \citep{2015A&A...582A..49H,2014A&A...564A.133J,2015A&A...582A..81J}, but since there are only 4 stars in this sample observed with APOGEE that have all three parameters calibrated in DR14, this sample was also not used. 

342 of the 1304 stars analyzed by \citet{2017AJ....154..107P} are among the APOGEE targets, but only 20 of them are subgiants and in the regime with calibrated surface gravities.

Based on these lines of reasoning, the five comparison samples we have found most relevant for our purposes are  the samples of \citet{2016ApJS..225...32B}, \citet{2015A&A...580A..24D}, the Gaia-ESO DR3, the sample of \citet{2017A&A...598A.100J}, and our own sample based on analysis of optical spectra from the ARCES spectrometer. They are all described in more detail in Sections \ref{sec:bacchus}-\ref{sec:j17} below, and their HR-diagrams are shown in Figure \ref{fig:hr-used}. All these analyses have been made with 1D LTE models, except in the case of iron, for which (very small) NLTE-corrections have been applied in \citet{2017A&A...598A.100J}.

\begin{figure*}
\epsscale{1.18}
\plotone{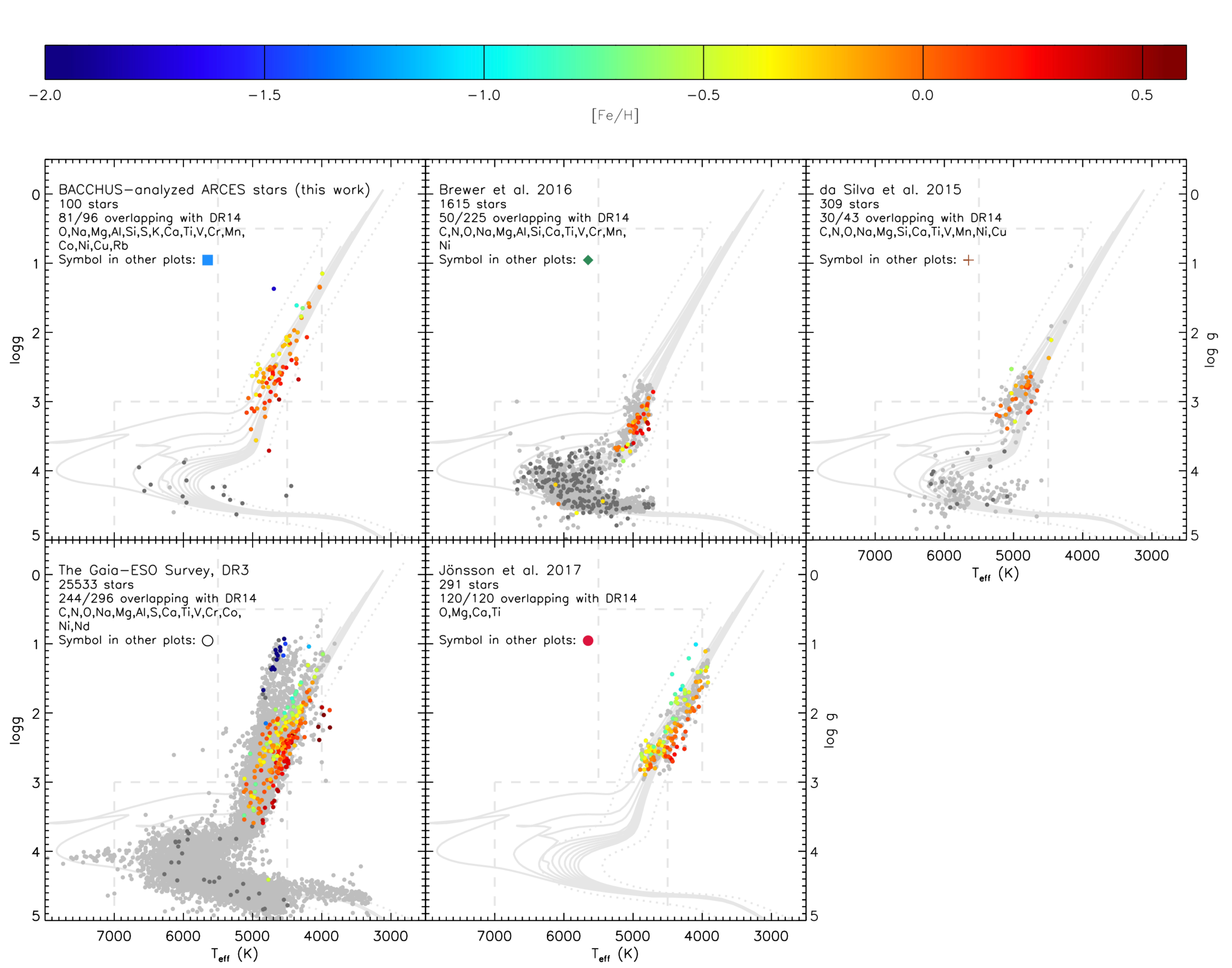}
\caption{HR diagrams for the reference samples used in the comparisons in Sections \ref{sec:params}-\ref{sec:conclusions}. In the comparisons made in this paper, only stars with calibrated values for all three stellar parameters in APOGEE are used, and these stars  (for DR14) are plotted color-coded in metallicity, the stars that are not in DR14 are plotted in gray, and the stars that do overlap with DR14 but only have uncalibrated values for one or more stellar parameter are plotted in darker gray. The elements overlapping between APOGEE and the reference in question is listed in each panel. As a guide for the eye, the same isochrones and regions as in Figure \ref{fig:hr-apogee} are plotted in every panel.\label{fig:hr-used}}
\end{figure*}

\subsubsection{BACCHUS analyzed ARCES-stars}\label{sec:bacchus}
We have observed a sample of 100 stars using the optical spectrometer ARCES (R$\sim$32,000) on the Apache Point 3.5m telescope. The stars were chosen from the APOGEE catalogue to have a spread in stellar parameters, and include both dwarfs and giants with a wide range of metallicities. The stars have $0.0<V<11.1$ and the spectra have S/N that ranges from $50 \leq S/N \leq 300$, with a median S/N of 115 around 6000~\AA. 

For determination of the stellar parameters as well as the abundances of O, Na, Mg, Al, Si, S, K, Ca, Ti, V, Cr, Mn, Co, Ni, Cu, Rb, and Y, we used the Brussels Automatic Code for Characterizing High AccUracy Spectra \citep[henceforth BACCHUS;][]{2016ascl.soft05004M}. BACCHUS is a stellar parameter and abundance analysis pipeline that uses Turbospectrum in combination with MARCS spherical 1D LTE models. The model atmosphere grid is alpha-enhanced for the lower metallicities according to the `standard' MARCS-scheme. The stellar parameters are determined in the classical way, demanding excitation and ionization equilibrium using a set of Fe I and Fe II lines. The analysis performed is similar to that described in \citet{2015MNRAS.447.2046H}, with the exception of the line list used: here we used the Gaia-ESO line list \citep[v.5,][Heiter et al. in prep.]{2015PhyS...90e4010H}, complemented with line information from the VALD database \citep{2000BaltA...9..590K,2015PhyS...90e4005R} for the non-covered wavelength-regimes in the Gaia-ESO list.

 The performance of BACCHUS has been thoroughly tested against a set of well-known Gaia benchmark stars \citep[e.g.][]{2014A&A...564A.133J,2015A&A...582A..81J,2015PhyS...90e4010H,2016A&A...592A..70H} and found to be both accurate and precise. One particular strength of BACCHUS is that it uses spherical radiative transfer in the spectral synthesis, something that is recommended when analyzing giants \citep{2006A&A...452.1039H}. We refer the reader to Section 4.3.3 of \citet{2014A&A...564A.133J} and Section 2.2 of \citet{2015MNRAS.447.2046H} for more details about BACCHUS and how it is employed for stellar parameters and abundance analysis.

Three tables present the results from this analysis: Table \ref{tab:linedata} presents the atomic data used, Table \ref{tab:linetoline} presents abundances from individual spectral lines, and Table \ref{tab:bacchus} summarizes the adopted stellar parameters and abundances for each star. These tables are given in their entirety in the electronic version.

\begin{deluxetable*}{lrrr}
\tablecaption{The line data used in the BACCHUS analysis. This is only an excerpt of the table to show its form and content. The complete table is available in electronic form.\label{tab:linedata}}
\tablehead{
\colhead{Element} & \colhead{Wavelength (\AA)} & \colhead{$\log gf$} & \colhead{$E_{\mathrm{low}}$ (eV)}
}
\startdata
[O I]  &   6300.3038 &  -9.7150  &  0.0000\\   
Na I   &   5682.6333 &  -0.7060  &  2.1020\\   
Mg I   &   5711.0880 &  -1.7240  &  4.3460\\   
Mg I   &   8712.6890 &  -1.2130  &  5.9320\\   
Mg I   &   8717.8250 &  -0.8660  &  5.9330\\   
...    &         ... &      ...  &     ...\\   
\enddata    
\end{deluxetable*}

\begin{deluxetable*}{lccc}
\tablecaption{The line-to-line abundances from the BACCHUS analysis. This is only an excerpt of the table to show its form and content. The complete table is available in electronic form.\label{tab:linetoline}}
\tablehead{
\colhead{Star} & \colhead{Element} & \colhead{Line (\AA)} & \colhead{Abundance}
}
\startdata
2MASSJ00002012+5612368 &  Al  &  5557.1 &  6.62\\
2MASSJ00002012+5612368 &  Al  &  6696.0 &  6.68\\
2MASSJ00002012+5612368 &  Al  &  6698.7 &  6.76\\
2MASSJ00002012+5612368 &  Al  &  7835.3 &  6.72\\
2MASSJ00002012+5612368 &  Al  &  7836.1 &  6.70\\
                   ... & ...  &     ... &   ...\\
\enddata    
\end{deluxetable*}

\begin{deluxetable*}{ccccccccc}
\tablecaption{The stellar parameters and abundances from the BACCHUS analysis. All abundances are relative to the solar abundances of \citet{2007SSRv..130..105G}. This is only an excerpt of the table to show its form and content. The complete table is available in electronic form.\label{tab:bacchus}}
\tablehead{
\colhead{Star} & \colhead{$T_{\mathrm{eff}}$} & \colhead{$\log g$} & \colhead{[Fe/H]} & \colhead{$v_{\mathrm{mic}}$} & \colhead{[O/Fe]} & \colhead{[Na/Fe]} & \colhead{[Mg/Fe]} & \colhead{...}
}
\startdata
2MASSJ00002012+5612368 & 4751 $\pm$  75 & 2.67 $\pm$ 0.49 &  0.22 $\pm$ 0.09 & 1.18 $\pm$ 0.07 & -0.20 $\pm$ 0.05 &  0.04 $\pm$ 0.08 & -0.06 $\pm$ 0.03 & ...\\
2MASSJ00012723+8520108 & 5956 $\pm$  16 & 4.14 $\pm$ 0.06 &  0.20 $\pm$ 0.07 & 1.15 $\pm$ 0.04 &   ... $\pm$  ... &  0.19 $\pm$ 0.04 &  0.06 $\pm$ 0.03 & ...\\
2MASSJ00041502+5614532 & 4596 $\pm$  57 & 2.74 $\pm$ 0.12 &  0.24 $\pm$ 0.09 & 0.97 $\pm$ 0.07 &   ... $\pm$  ... & -0.07 $\pm$ 0.07 & -0.17 $\pm$ 0.03 & ...\\
2MASSJ00100473+8601230 & 6565 $\pm$  13 & 4.29 $\pm$ 0.29 &  0.06 $\pm$ 0.12 & 1.18 $\pm$ 0.07 &   ... $\pm$  ... &  0.14 $\pm$ 0.12 &  0.08 $\pm$ 0.07 & ...\\
2MASSJ00202846+6238519 & 4825 $\pm$  27 & 3.02 $\pm$ 0.29 &  0.17 $\pm$ 0.08 & 1.16 $\pm$ 0.05 &   ... $\pm$  ... &  0.05 $\pm$ 0.04 & -0.07 $\pm$ 0.02 & ...\\
... & ... & ... & ... & ... & ... & ... & ... & ...\\ 
\enddata    
\end{deluxetable*}

The elements that overlap with DR13/DR14 are O, Na, Mg, Al, Si, S, K, Ca, Ti, V, Cr, Mn, Co, Ni, Cu, and Rb and the number of non-flagged stars overlapping with DR13 are 83/98 (here and subsequently, these two numbers give the number of stars with all calibrated parameters / number with at least one uncalibrated parameter), and the number of non-flagged stars overlapping with DR14 are 81/96.

\subsubsection{Brewer et al. (2016)}
\citet{2016ApJS..225...32B} determined the abundances of C, N, O, Na, Mg, Al, Si, Ca, Ti, V, Cr, Mn, Ni, and Y in a sample of 1615 stars with $0.0<V<16.4$. The optical spectra used were recorded using the spectrometer HIRES (R$\sim$70,000) at the 10m telescope at Keck Observatory. They find that their precision decreased significantly for the 424 stars with S/N$<100$ compared to the 1191 stars with S/N$>100$. 

The elements that overlap between this study and DR13/DR14 are C, N, O, Na, Mg, Al, Si, Ca, Ti, V, Cr, Mn, and Ni, and the number of non-flagged stars overlapping with DR13 are 75/187 and the number of non-flagged stars overlapping with DR14 are 50/225.

The stellar parameters were spectroscopically determined using the $\chi^2$-minimizing spectral synthesis code Spectroscopy Made Easy \citep[SME;][]{1996A&AS..118..595V} on several spectral features, for example Fe I and Fe II lines, and  the wings of the Mg I b triplet. The same code was used to determine the stellar abundances. They use ATLAS plane-parallel model atmospheres, and the line list used was originally from VALD, but the wavelengths, transition probabilities, and van der Waals broadening were changed to fit a spectrum of the Sun.

\subsubsection{da Silva et al. (2015)}
\citet{2015A&A...580A..24D} determined the abundances of C, N, O, Na, Mg, Si, Ca, Ti, V, Mn, Ni, Cu, and Ba in a sample of 309 stars with $1.1<V<9.7$. The optical spectra were recorded using the spectrometer ELODIE (R$\sim$42,000) at the Haute Provence Observatory. A subsample of 172 of the stars were previously analyzed in \citet{2011A&A...526A..71D}, but in the 2015 paper the sample was expanded, and more elemental abundances were determined.

The elements that overlap between this study and DR13/DR14 are C, N, O, Na, Mg, Si, Ca, Ti, V, Mn, Ni, and Cu and the number of non-flagged stars overlapping with DR13 are 33/38 and the number of non-flagged stars overlapping with DR14 are 30/43.

The stellar parameters were determined using the automatic equivalent width measurement code ARES \citep{2007A&A...469..783S} and MOOG on Fe I and Fe II lines. ATLAS plane-parallel model atmospheres were used. The same method was used when determining the elemental abundances, except in the cases of C, N, O, and Na, where the spectral synthesis mode in MOOG was used instead of the equivalent width method. The original line list used was taken from VALD, but then it was astrophysically calibrated to fit the solar spectrum.

\subsubsection{The Gaia-ESO survey}
The Gaia-ESO survey \citep{2012Msngr.147...25G} is an ongoing optical spectroscopic survey that so far has observed more than 83,000 stars in which they intend to determine abundances of Li, C, N, O, Na, Mg, Al, S, Ca, Sc, Ti, V, Cr, Co, Ni, Zn, Y, Zr, Ba, La, Ce, Nd, and Eu. In their latest data release (DR3, from May 2017), they present stellar parameters and abundances of 25,533 stars\footnote{\url{http://www.eso.org/rm/api/v1/public/releaseDescriptions/92}}. The spectrometers FLAMES-GIRAFFE (R$\sim$20,000) and FLAMES-UVES (R$\sim$47,000) at the Very Large telescope (VLT) are used for carrying out the observations, and the targeted stars have $12\le J \le 17.5$. 

The elements that overlap between DR3 of Gaia-ESO and DR13/DR14 are C, N, O, Na, Mg, Al, S, Ca, Ti, V, Cr, Co, Ni, and Nd, and the number of non-flagged stars overlapping with DR13 are 139/152 and the number of non-flagged stars overlapping with DR14 are 244/278. 

The analysis of the Gaia-ESO spectra are done by several research groups -- nodes -- using their own preferred method. Some nodes use equivalent width methods and others spectral synthesis. In the end, all these results are averaged using an elaborate scheme based on the performance of the particular nodes for different types of stars \citep{2014A&A...570A.122S}. To try to minimize the systematic differences between the nodes, they all use the same model atmospheres (spherical MARCS for giants, and plane parallel MARCS for dwarfs) and line list. A significant amount of work has been devoted to finding and vetting atomic data \citep[][Heiter et al. in prep.]{2015PhyS...90e4010H}, a task that has benefitted many independent optical stellar spectroscopic works.

\subsubsection{J\"onsson et al. (2017)}\label{sec:j17}
\citet{2017A&A...598A.100J} determined the abundances of O, Mg, Ca, and Ti in a sample of 291 stars with $0.0<V<11.9$.  The optical spectra used were recorded with the spectrometers FIES (R$\sim$67,000) at the Nordic Optical telescope, NARVAL (R$\sim$65,000) at the T\'elescope Bernard Lyot, and ESPaDOnS (R$\sim$65,000) at the Canada-France-Hawaii Telescope. The spectra have $30<S/N<250$, but most close to 100.

The elements that overlap between this study and DR13/DR14 are O, Mg, Ca, and Ti and the number of non-flagged stars overlapping with DR13 are 106/106 and the number of non-flagged stars overlapping with DR14 are 120/120.

The stellar parameters were determined using the SME code on Fe I, Fe II lines and wings of strong Ca I lines. The analysis was made using MARCS spherical 1D LTE models that were alpha-enhanced for the lower metallicities according to the `standard' MARCS-scheme, and a slightly updated version of the Gaia-ESO line list (v.5, see their Section 3.1). NLTE-corrections for Fe were applied \citep{2012MNRAS.427...50L}.
  
\subsection{Globular cluster star samples}
Multiple populations in GCs are extensively studied in the literature using both photometric and spectroscopic data. To date, almost all GCs have been found to have multiple main sequences and/or subgiant and/or giant branches, \citep[e.g.,][]{2007ApJ...661L..53P,2008ApJ...673..241M,2015AJ....149...91P}. These different populations in metal-poor clusters have different chemical compositions. For example, sodium and oxygen, are found to vary such that the stars that were formed first -- the first generation (FG) stars -- are sodium poor and oxygen rich, while the SG stars are sodium rich and oxygen poor \citep[e.g.,][]{2009A&A...508..695C,2009A&A...505..139C,2009A&A...505..117C}.

As mentioned in \citet{2015AJ....150..148H}, the ASPCAP-team has long suspected that the pipeline is not performing optimally for stars with extreme types of `non-standard' elemental abundance patterns, like SG GC stars. In Section \ref{sec:sg}, we attempt to quantize these problems by comparing APOGEE DR13, DR14, and the Cannon results to those of independent analyses of cluster stars.

\citet{2015AJ....149..153M} analyzed 428 giant stars in 10 northern GC using actual APOGEE DR10 spectra, photometric effective temperatures, and surface gravities determined by isochrone fitting. They also made an extensive cross-match between the APOGEE-observed GC-stars and previous works \citep[in this paper we use][]{2009A&A...505..139C,2000AJ....120.1364C,2005AJ....129..303C,2001AJ....122.1438I,2005PASP..117.1308J,2012ApJ...754L..38J,2010AJ....139.2289K,1992AJ....104..645K,2003PASP..115..143K,2011AJ....141...62L,1996ApJ...470..953M,2011PASP..123.1139O,2003AJ....125..224R,1996AJ....112.1517S,1991AJ....102.2001S,1992AJ....104.2121S,1997AJ....114.1964S,2000AJ....120.1351S,2004AJ....127.2162S,2006ApJ...638.1018Y,2008ApJ...689.1020Y}. We have chosen to compare the APOGEE-results to those of the optical references in \citet{2015AJ....149..153M}, and not the actual results in \citet{2015AJ....149..153M} to be consistent with the rest of the paper, where optical reference works are used. However, we have checked and found that the conclusions would remain the same if the values from the independent H-band analysis of \citet{2015AJ....149..153M} are used to compare to the APOGEE abundances.

\section{Comparing the stellar parameters}
\subsection{Field star samples}\label{sec:params}
\begin{figure*}
\epsscale{1.23}
\plotone{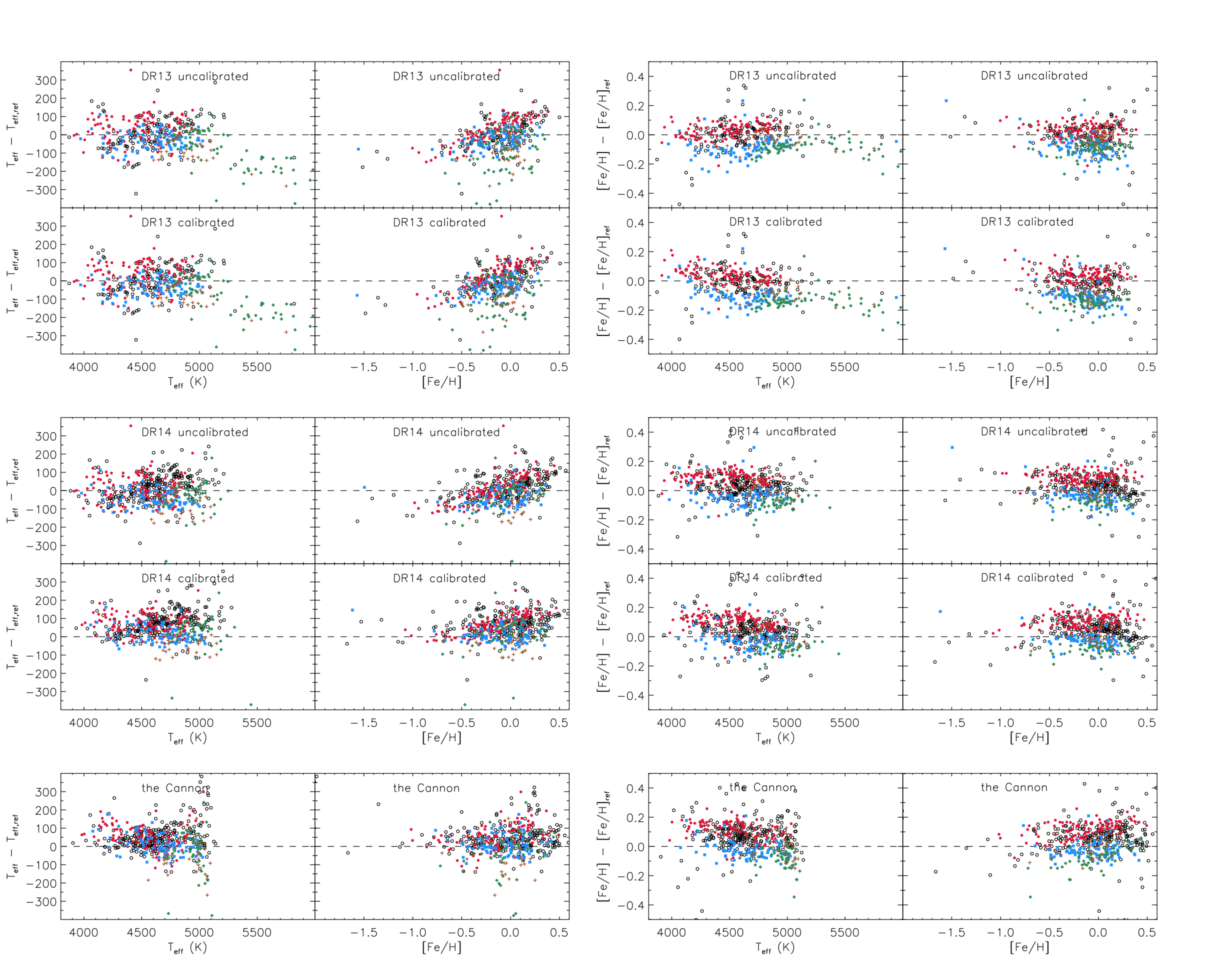}
\caption{Differences in T$_\mathrm{eff}$ and [Fe/H] for DR13, DR14, the Cannon, and the references. The BACCHUS analyzed ARCES-stars are marked using blue squares, the \citet{2016ApJS..225...32B}-stars are marked using green diamonds, the \citet{2015A&A...580A..24D}-stars are marked using brown crosses, the values from Gaia-ESO DR3 are marked using black open circles, and the \citet{2017A&A...598A.100J}-stars are marked using red dots.\label{fig:deltaplots}}
\end{figure*}

Figure \ref{fig:deltaplots} shows the comparison between the ASPCAP DR13, DR14, and the Cannon parameters and those of the reference samples, as a function of both effective temperature and metallicity. Table \ref{tab:params} summarizes the mean differences and scatter. We find, in most cases, scatter that is consistent with the combined uncertainties of the samples being compared.

\begin{deluxetable*}{llcccccc}
\tablecaption{Median and standard deviation for the differences in stellar parameters between APOGEE and the references, in the sense APOGEE - reference. Note that these values encompass systematic and random uncertainties in the APOGEE analysis as well as the reference work. The number in parenthesis is the number of stars used in the comparison. The values for effective temperatures are not representative, since there is a trend of effective temperature in the APOGEE data. For more information, see Section \ref{sec:params}.\label{tab:params}}
\tablehead{
\colhead{ } & \colhead{ } & \colhead{BACCHUS} & \colhead{Brewer+(2016)} & \colhead{da Silva+(2015)}& \colhead{Gaia-ESO DR3} & \colhead{J\"onsson+(2017)}& \colhead{All}
}
\startdata
T$_{\mathrm{eff}}$ & DR13 &-31 $\pm$ 79 (83) & -62 $\pm$ 157 (75) & -84 $\pm$ 85 (33) & 10 $\pm$ 106 (139) & 34 $\pm$ 75 (106) & -14 $\pm$ 115 (436)\\
   & DR14 &3 $\pm$ 45 (81) & 34 $\pm$ 259 (50) & -35 $\pm$ 47 (30) & 72 $\pm$ 70 (244) & 67 $\pm$ 61 (120) & 53 $\pm$ 108 (525)\\
   & Cannon &4 $\pm$ 53 (79) & -26 $\pm$ 303 (55) & -67 $\pm$ 87 (27) & 40 $\pm$ 76 (248) & 53 $\pm$ 74 (108) & 30 $\pm$ 132 (517)\\
\hline
$\log g$ & DR13 &-0.10 $\pm$ 0.19 (83) & -0.05 $\pm$ 0.21 (75) & -0.23 $\pm$ 0.16 (33) & -0.00 $\pm$ 0.22 (139) & -0.04 $\pm$ 0.12 (106) & -0.06 $\pm$ 0.20 (436)\\
   & DR14 &-0.13 $\pm$ 0.21 (81) & -0.08 $\pm$ 0.26 (50) & -0.28 $\pm$ 0.14 (30) & -0.03 $\pm$ 0.20 (244) & -0.05 $\pm$ 0.13 (120) & -0.08 $\pm$ 0.20 (525)\\
   & Cannon &-0.15 $\pm$ 0.17 (79) & -0.12 $\pm$ 0.25 (55) & -0.18 $\pm$ 0.14 (27) & -0.02 $\pm$ 0.22 (248) & -0.04 $\pm$ 0.14 (108) & -0.07 $\pm$ 0.21 (517)\\
\hline
[Fe/H] & DR13 &-0.11 $\pm$ 0.08 (83) & -0.14 $\pm$ 0.06 (75) & -0.07 $\pm$ 0.05 (33) & -0.00 $\pm$ 0.13 (139) & 0.04 $\pm$ 0.05 (106) & -0.04 $\pm$ 0.11 (436)\\
   & DR14 &-0.03 $\pm$ 0.07 (81) & -0.07 $\pm$ 0.06 (50) & 0.01 $\pm$ 0.06 (30) & 0.05 $\pm$ 0.12 (244) & 0.11 $\pm$ 0.05 (120) & 0.04 $\pm$ 0.10 (525)\\
   & Cannon &-0.04 $\pm$ 0.06 (79) & -0.07 $\pm$ 0.09 (55) & 0.01 $\pm$ 0.08 (27) & 0.07 $\pm$ 0.11 (248) & 0.12 $\pm$ 0.06 (108) & 0.05 $\pm$ 0.11 (517)\\
\enddata    
\end{deluxetable*}

In DR14, calibrated surface gravities were not provided for warmer stars, hence these stars appear in the panels for DR13 but not for DR14.

The DR14 calibrated values and the values as determined from the Cannon are expected to be similar, since the Cannon has been trained on DR14 calibrated results, which is the case, both when looking at Figure \ref{fig:deltaplots} and Table \ref{tab:params}. The calibrated DR13 effective temperatures show a systematic offset of only -14 K and scatter of 115 K when compared to the references, for DR14 the same values are +53 K and 108 K, and for the DR14 Cannon-analysis the values are +30 K and 132 K, indicating a slightly higher scatter for the Cannon effective temperatures. From the bottom panel in the leftmost column of plots, it is apparent that the scatter is increasing in the Cannon-analysis for effective temperatures around 5000 K. This can possibly be traced to the upwards `flare' left of the red clump in the Cannon HR-diagram in Figure \ref{fig:hr-apogee}. 

In the second column of panels in Figure \ref{fig:deltaplots}, a more or less clear trend is seen in the effective temperature difference between DR13 and all comparison samples as a function of [Fe/H]. This trend was also found when comparing the ASPCAP effective temperatures to effective temperatures derived from photometry. While no effective temperature calibration was applied in DR13, \citet{paperi} suggests a relation to be applied to DR13 to remove this effect. A similar effective temperature correction was applied as part of the DR14 calibrations, and consequently, the trend is much less pronounced in the plot showing the DR14 calibrated parameters (the fourth row, second column panel in  Figure \ref{fig:deltaplots}). However, a weak residual is still present in DR14 (and the DR14 Cannon-values) for the most metal-rich stars, in the sense that the calibrated DR14 ASPCAP effective temperatures are approximately 100 K higher than the optical effective temperatures. 

Even if the ASPCAP-trend of effective temperature with metallicity is reduced with the calibrated effective temperatures presented in DR14, any trend can potentially have far-reaching consequences for the derived abundances, since the uncalibrated stellar parameters (effective temperature, surface gravity, and metallicity) are used when determining the abundances in ASPCAP. This methodology is motivated by the fact that in H-band spectra of giants at the resolution of APOGEE, many of the spectral lines of interest are somewhat blended by the vast amount of molecular lines present, and using the stellar parameters that -- on a global level -- best fits the spectrum will do the best job at synthesizing, and hence removing the impact of, blending lines. However, some elements whose spectral lines show a large dependence on the adopted effective temperature might be more precisely determined if calibrated stellar parameters were to be used instead (see discussion in Sections \ref{sec:c}-\ref{sec:nc}).

In the two rightmost columns of panels in Figure \ref{fig:deltaplots} and in Table \ref{tab:params}, systematic zero-point differences regarding metallicity-scales can be seen: for example, DR13 is systematically $\sim0.04$ dex higher in [Fe/H] than what is found in \citet{2017A&A...598A.100J} for the very same stars, while $\sim0.11$ dex lower than the BACCHUS analyzed ARCES-stars. [Fe/H] in DR14 is $\sim0.11$ dex higher than \citet{2017A&A...598A.100J}, and $\sim0.03$ dex lower than the BACCHUS analyzed ARCES-stars. There is a possible negative trend for the $\Delta$[Fe/H] as a function of T$_\mathrm{eff}$ for the calibrated DR13-values (surprisingly not as obvious for the DR13 uncalibrated values), but in no other panels in the two rightmost columns can obvious trends be seen when all reference values are taken into account. The fact that different reference samples do not always agree with each other highlights the challenges of determining accurate stellar parameters.

\subsection{Globular cluster star samples}\label{sec:sg}
\begin{figure*}
\epsscale{1.18}
\plotone{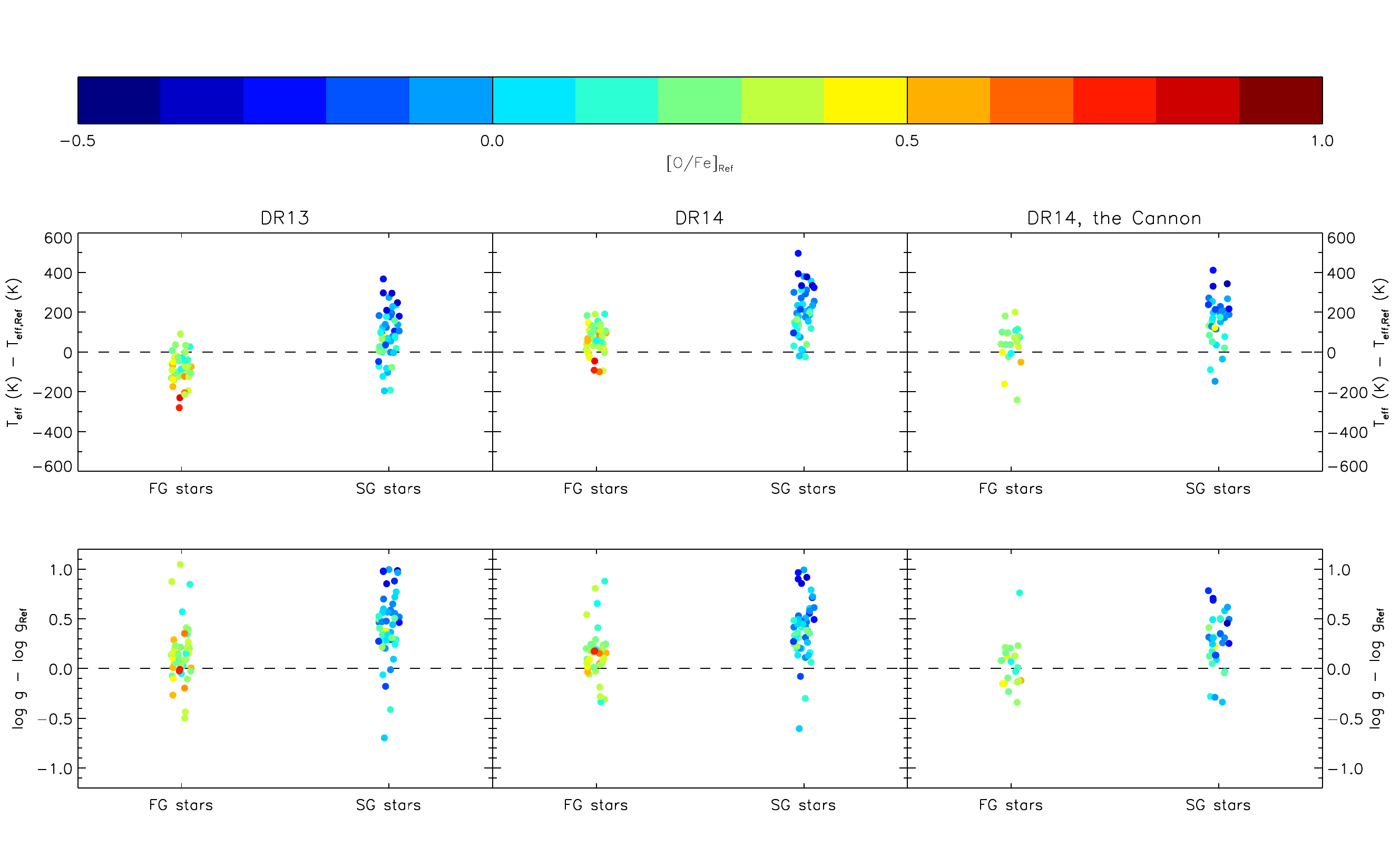}
\caption{Differences of effective temperature and surface gravities as determined by DR13, DR14, and the Cannon as compared to the optical references for first and second generation globular cluster stars, color-coded with [O/Fe] from the reference in question. The stars have been randomly spread out somewhat along the x-axis to make the plot clearer and show all points.\label{fig:sg}}
\end{figure*}

In Figure \ref{fig:sg} we have plotted differences of effective temperature and surface gravities as determined by DR13, DR14, and the Cannon as compared to the optical references for FG and SG GC stars. 

From this plot, one can draw the conclusion that the APOGEE analyses are over-estimating (with respect to the optical studies taken as references) effective temperatures and surface gravities for SG stars, and especially so for the stars with extreme SG-type abundance pattern (the oxygen-poor, blue points). For FG stars, the spread in effective temperatures are about the same as when comparing to the disk-type abundance-pattern stars in Figure \ref{fig:deltaplots}.

It is also clear that the APOGEE stellar parameters for the extreme SG-type field star from \citet{2016ApJ...833..132F} actually is \emph{expected} to show the large deviations found in \citet{2017MNRAS.469..774P} ($\Delta$T$_{\mathrm{eff}}=$+317 K and $\Delta\log g=$+0.89 in the sense APOGEE - \citet{2017MNRAS.469..774P}). These large inaccuracies in stellar parameters obviously heavily influence all the APOGEE-determined abundances for the extreme SG-type stars. This is also easily confirmed from comparing to  \citet{2017MNRAS.469..774P} where abundance differences of up to 0.5 dex compared to DR14 can be seen. Unfortunately, all inaccuracies in the APOGEE abundances seem to work in the sense that they tend to erase the SG abundance pattern of C, N, O, Na, and Al of the star, i.e., compared to the references, the SG-typical low carbon and oxygen abundances are determined higher by ASPCAP, and vice versa for the SG-typical high nitrogen, sodium, and aluminum abundances that are determined lower by ASPCAP.

In the rest of this paper we avoid using comparison works with an expected high ratio of `non-standard' abundance-pattern stars. However, from the design and target selection of the APOGEE survey, the number of stars with these types of extreme SG abundance patterns are believed to be very small. For example, \citet{2017ApJ...846L...2F} conducted a search for such stars within DR13 and found 260 stars. However, as described above, ASPCAP analysis of stars with atypical abundance patterns leads to systematic errors in the stellar parameters that in turn result in systematic errors in chemical abundances such that these stars appear less atypical, so the actual number of stars in the DR13 sample with SG-type abundance pattern might be much higher.

\section{Comparing the abundances}\label{sec:abunds}
In this section, we assess the APOGEE/ASPCAP chemical abundances element by element. Table \ref{tab:abunds} presents a summary, showing the median difference and spread between the ASPCAP analysis and the literature values for elements that, as discussed below, do not show systematic trends with any stellar parameter. Table \ref{tab:trendabunds} shows comparable results for other elements, although these may be less meaningful because there are systematic trends for these. Generally speaking, one can see from Table \ref{tab:abunds} that DR14 appear to have more precise and accurate abundances than DR13. The spread of the Cannon abundances is greater than the spread of ASPCAP DR14 abundances for all elements but silicon. It is important to note that while, for example, there are 50 stars in \citet{2016ApJS..225...32B} that have calibrated stellar abundances in DR14, there are not necessarily 50 Brewer-points in every element comparison: for some stars either \citet{2016ApJS..225...32B} or ASPCAP might have failed to determine the abundance in question. The same is true for all comparison samples and elements. This and the fact that not all APOGEE abundances are derived in all comparison works, results in some elements being more thoroughly evaluated than others, something that can bee seen in the figures in this Section as well as in Tables \ref{tab:abunds}-\ref{tab:trendabunds}.

\begin{deluxetable*}{llcccccc}
\tablecaption{Median and standard deviation for the differences in abundances between APOGEE and the references, in the sense APOGEE - reference, for the elements not showing any trends with stellar parameters when comparing to the references. Note that these values encompass systematic and random uncertainties in as well the APOGEE analysis and the reference work. The number in parenthesis is the number of stars used in the comparison. None of the reference works have determined phosphorous abundances.  For more information, see the relevant Sections \ref{sec:c}-\ref{sec:nc}.\label{tab:abunds}}
\tablehead{
\colhead{ } & \colhead{ } & \colhead{BACCHUS} & \colhead{Brewer+(2016)} & \colhead{da Silva+(2015)} & \colhead{Gaia-ESO DR3} & \colhead{J\"onsson+(2017)} & \colhead{All}
}
\startdata
C  & DR13 &... & -0.04 $\pm$ 0.09 (32) & -0.12 $\pm$ 0.06 (23) & 0.12 $\pm$ 0.12 (35) & ... & -0.02 $\pm$ 0.13 (90)\\
   & DR14 &... & 0.00 $\pm$ 0.12 (34) & -0.08 $\pm$ 0.06 (24) & 0.15 $\pm$ 0.12 (37) & ... & 0.01 $\pm$ 0.15 (95)\\
   & Cannon &... & 0.03 $\pm$ 0.12 (54) & -0.09 $\pm$ 0.08 (27) & 0.07 $\pm$ 0.45 (48) & ... & -0.00 $\pm$ 0.29 (129)\\
\hline
C I & DR13 &... & -0.09 $\pm$ 0.08 (36) & -0.18 $\pm$ 0.07 (23) & 0.07 $\pm$ 0.12 (35) & ... & -0.10 $\pm$ 0.14 (94)\\
   & DR14 &... & -0.06 $\pm$ 0.11 (35) & -0.16 $\pm$ 0.08 (25) & 0.06 $\pm$ 0.13 (39) & ... & -0.06 $\pm$ 0.15 (99)\\
   & Cannon &... & 0.00 $\pm$ 0.14 (54) & -0.18 $\pm$ 0.08 (27) & 0.02 $\pm$ 0.32 (48) & ... & -0.03 $\pm$ 0.22 (129)\\
\hline
Na & DR13 &-0.15 $\pm$ 0.12 (73) & -0.13 $\pm$ 0.09 (34) & -0.26 $\pm$ 0.28 (21) & -0.27 $\pm$ 0.26 (34) & ... & -0.18 $\pm$ 0.19 (162)\\
   & DR14 &-0.00 $\pm$ 0.11 (70) & -0.04 $\pm$ 0.10 (25) & -0.14 $\pm$ 0.21 (20) & -0.10 $\pm$ 0.22 (29) & ... & -0.03 $\pm$ 0.16 (144)\\
   & Cannon &0.01 $\pm$ 0.17 (78) & 0.01 $\pm$ 0.29 (55) & -0.07 $\pm$ 0.16 (25) & -0.07 $\pm$ 0.31 (41) & ... & -0.02 $\pm$ 0.24 (199)\\
\hline
Mg & DR13 &-0.11 $\pm$ 0.07 (79) & -0.04 $\pm$ 0.06 (36) & -0.16 $\pm$ 0.06 (23) & -0.12 $\pm$ 0.12 (114) & -0.05 $\pm$ 0.07 (105) & -0.09 $\pm$ 0.10 (357)\\
   & DR14 &-0.03 $\pm$ 0.08 (79) & 0.04 $\pm$ 0.06 (49) & -0.08 $\pm$ 0.07 (30) & -0.02 $\pm$ 0.10 (222) & 0.02 $\pm$ 0.08 (117) & -0.01 $\pm$ 0.09 (497)\\
   & Cannon &-0.03 $\pm$ 0.08 (79) & 0.02 $\pm$ 0.11 (55) & -0.11 $\pm$ 0.07 (27) & 0.00 $\pm$ 0.16 (244) & 0.00 $\pm$ 0.09 (108) & -0.01 $\pm$ 0.13 (513)\\
\hline
Al & DR13 &-0.18 $\pm$ 0.06 (78) & -0.04 $\pm$ 0.25 (36) & ... & -0.09 $\pm$ 0.15 (111) & ... & -0.12 $\pm$ 0.15 (225)\\
   & DR14 &-0.06 $\pm$ 0.07 (78) & 0.05 $\pm$ 0.12 (34) & ... & 0.04 $\pm$ 0.16 (172) & ... & 0.01 $\pm$ 0.14 (284)\\
   & Cannon &-0.04 $\pm$ 0.09 (79) & -0.05 $\pm$ 0.16 (54) & ... & 0.03 $\pm$ 0.26 (234) & ... & -0.01 $\pm$ 0.22 (367)\\
\hline
Si & DR13 &-0.13 $\pm$ 0.08 (79) & 0.01 $\pm$ 0.08 (36) & -0.15 $\pm$ 0.05 (23) & ... & ... & -0.12 $\pm$ 0.10 (138)\\
   & DR14 &-0.04 $\pm$ 0.08 (79) & 0.06 $\pm$ 0.14 (49) & -0.07 $\pm$ 0.05 (30) & ... & ... & -0.03 $\pm$ 0.11 (158)\\
   & Cannon &-0.04 $\pm$ 0.09 (79) & 0.07 $\pm$ 0.12 (55) & -0.07 $\pm$ 0.05 (27) & ... & ... & -0.03 $\pm$ 0.10 (161)\\
\hline
S  & DR13 &-0.08 $\pm$ 0.14 (65) & ... & ... & -0.03 $\pm$ 0.19 (33) & ... & -0.06 $\pm$ 0.16 (98)\\
   & DR14 &-0.06 $\pm$ 0.12 (66) & ... & ... & 0.00 $\pm$ 0.15 (37) & ... & -0.03 $\pm$ 0.13 (103)\\
   & Cannon &-0.05 $\pm$ 0.13 (65) & ... & ... & -0.06 $\pm$ 0.16 (37) & ... & -0.05 $\pm$ 0.14 (102)\\
\hline
Ca & DR13 &-0.18 $\pm$ 0.08 (79) & -0.12 $\pm$ 0.05 (36) & -0.07 $\pm$ 0.05 (22) & 0.04 $\pm$ 0.12 (84) & 0.01 $\pm$ 0.06 (105) & -0.04 $\pm$ 0.12 (326)\\
   & DR14 &-0.11 $\pm$ 0.08 (79) & -0.04 $\pm$ 0.13 (35) & -0.01 $\pm$ 0.07 (24) & 0.12 $\pm$ 0.14 (138) & 0.05 $\pm$ 0.07 (120) & 0.03 $\pm$ 0.14 (396)\\
   & Cannon &-0.10 $\pm$ 0.16 (79) & -0.05 $\pm$ 0.15 (55) & -0.05 $\pm$ 0.10 (26) & 0.14 $\pm$ 0.20 (173) & 0.04 $\pm$ 0.10 (108) & 0.01 $\pm$ 0.19 (441)\\
\hline
Ti II & DR13 &-0.11 $\pm$ 0.19 (78) & -0.13 $\pm$ 0.19 (36) & -0.11 $\pm$ 0.12 (22) & 0.06 $\pm$ 0.17 (107) & 0.04 $\pm$ 0.14 (101) & -0.03 $\pm$ 0.19 (344)\\
   & DR14 &-0.01 $\pm$ 0.16 (78) & -0.19 $\pm$ 0.21 (31) & -0.11 $\pm$ 0.16 (25) & 0.15 $\pm$ 0.22 (197) & 0.16 $\pm$ 0.14 (116) & 0.09 $\pm$ 0.21 (447)\\
   & Cannon &0.02 $\pm$ 0.20 (79) & -0.08 $\pm$ 0.25 (55) & -0.07 $\pm$ 0.28 (27) & 0.11 $\pm$ 0.35 (233) & 0.24 $\pm$ 0.23 (108) & 0.09 $\pm$ 0.31 (502)\\
\hline
Cr & DR13 &-0.05 $\pm$ 0.09 (79) & -0.10 $\pm$ 0.07 (36) & ... & 0.03 $\pm$ 0.12 (82) & ... & -0.03 $\pm$ 0.11 (197)\\
   & DR14 &0.02 $\pm$ 0.09 (80) & -0.03 $\pm$ 0.07 (35) & ... & 0.09 $\pm$ 0.14 (131) & ... & 0.04 $\pm$ 0.12 (246)\\
   & Cannon &0.01 $\pm$ 0.10 (79) & -0.06 $\pm$ 0.15 (55) & ... & 0.11 $\pm$ 0.28 (166) & ... & 0.04 $\pm$ 0.24 (300)\\
\hline
Mn & DR13 &0.03 $\pm$ 0.11 (78) & -0.21 $\pm$ 0.07 (36) & -0.05 $\pm$ 0.06 (23) & ... & ... & -0.03 $\pm$ 0.14 (137)\\
   & DR14 &0.10 $\pm$ 0.11 (79) & -0.11 $\pm$ 0.10 (35) & 0.03 $\pm$ 0.07 (25) & ... & ... & 0.05 $\pm$ 0.14 (139)\\
   & Cannon &0.09 $\pm$ 0.11 (79) & -0.11 $\pm$ 0.10 (55) & 0.01 $\pm$ 0.09 (27) & ... & ... & 0.01 $\pm$ 0.15 (161)\\
\hline
Ni & DR13 &-0.09 $\pm$ 0.10 (79) & -0.13 $\pm$ 0.05 (36) & -0.08 $\pm$ 0.06 (23) & 0.01 $\pm$ 0.09 (71) & ... & -0.07 $\pm$ 0.10 (209)\\
   & DR14 &0.01 $\pm$ 0.10 (79) & -0.05 $\pm$ 0.04 (35) & -0.00 $\pm$ 0.05 (25) & 0.09 $\pm$ 0.10 (104) & ... & 0.02 $\pm$ 0.10 (243)\\
   & Cannon &-0.01 $\pm$ 0.10 (79) & -0.05 $\pm$ 0.09 (55) & -0.02 $\pm$ 0.07 (27) & 0.10 $\pm$ 0.15 (131) & ... & 0.03 $\pm$ 0.14 (292)\\
\enddata    
\end{deluxetable*}

\begin{deluxetable*}{llcccccc}
\tablecaption{Median and standard deviation for the differences in abundances between APOGEE and the references, in the sense APOGEE - reference, for the elements showing trends with stellar parameters when comparing to the references that warrant further investigation. Note that these values encompass systematic and random uncertainties in as well the APOGEE analysis and the reference work. The number in parenthesis is the number of stars used in the comparison. For more information, see the relevant Sections \ref{sec:c}-\ref{sec:nc}.\label{tab:trendabunds}}
\tablehead{
\colhead{ } & \colhead{ } & \colhead{BACCHUS} & \colhead{Brewer+(2016)} & \colhead{da Silva+(2015)} & \colhead{Gaia-ESO DR3} & \colhead{J\"onsson+(2017)} & \colhead{All}
}
\startdata
N  & DR13 &... & 0.09 $\pm$ 0.15 (33) & -0.20 $\pm$ 0.08 (22) & 0.07 $\pm$ 0.17 (16) & ... & -0.03 $\pm$ 0.18 (71)\\
   & DR14 &... & 0.19 $\pm$ 0.19 (34) & -0.11 $\pm$ 0.09 (23) & 0.13 $\pm$ 0.15 (19) & ... & 0.08 $\pm$ 0.19 (76)\\
   & Cannon &... & 0.03 $\pm$ 0.21 (54) & -0.13 $\pm$ 0.13 (25) & 0.21 $\pm$ 0.23 (19) & ... & -0.00 $\pm$ 0.22 (98)\\
\hline
O  & DR13 &-0.30 $\pm$ 0.16 (51) & -0.02 $\pm$ 0.13 (34) & -0.38 $\pm$ 0.07 (23) & -0.06 $\pm$ 0.16 (36) & -0.05 $\pm$ 0.12 (75) & -0.12 $\pm$ 0.18 (219)\\
   & DR14 &-0.25 $\pm$ 0.16 (53) & 0.04 $\pm$ 0.15 (34) & -0.33 $\pm$ 0.07 (24) & -0.00 $\pm$ 0.17 (38) & -0.01 $\pm$ 0.13 (84) & -0.07 $\pm$ 0.19 (233)\\
   & Cannon &-0.22 $\pm$ 0.17 (54) & -0.03 $\pm$ 0.18 (54) & -0.36 $\pm$ 0.09 (27) & -0.03 $\pm$ 0.24 (49) & -0.04 $\pm$ 0.15 (78) & -0.09 $\pm$ 0.21 (262)\\
\hline
K  & DR13 &-0.30 $\pm$ 0.14 (57) & ... & ... & ... & ... & -0.30 $\pm$ 0.14 (57)\\
   & DR14 &-0.23 $\pm$ 0.15 (56) & ... & ... & ... & ... & -0.23 $\pm$ 0.15 (56)\\
   & Cannon &-0.18 $\pm$ 0.17 (57) & ... & ... & ... & ... & -0.18 $\pm$ 0.17 (57)\\
\hline
Ti I & DR13 &-0.16 $\pm$ 0.13 (79) & -0.06 $\pm$ 0.15 (36) & -0.07 $\pm$ 0.10 (23) & 0.03 $\pm$ 0.13 (109) & 0.02 $\pm$ 0.12 (105) & -0.03 $\pm$ 0.15 (352)\\
   & DR14 &-0.08 $\pm$ 0.13 (79) & 0.02 $\pm$ 0.14 (35) & -0.01 $\pm$ 0.10 (25) & 0.15 $\pm$ 0.13 (206) & 0.11 $\pm$ 0.12 (120) & 0.08 $\pm$ 0.15 (465)\\
   & Cannon &-0.06 $\pm$ 0.16 (79) & -0.06 $\pm$ 0.23 (55) & -0.01 $\pm$ 0.16 (27) & 0.12 $\pm$ 0.22 (233) & 0.07 $\pm$ 0.15 (108) & 0.05 $\pm$ 0.21 (502)\\
\hline
V  & DR13 &-0.03 $\pm$ 0.16 (76) & -0.04 $\pm$ 0.11 (34) & -0.11 $\pm$ 0.11 (23) & 0.03 $\pm$ 0.15 (34) & ... & -0.04 $\pm$ 0.14 (167)\\
   & DR14 &0.03 $\pm$ 0.15 (76) & 0.08 $\pm$ 0.23 (34) & 0.01 $\pm$ 0.12 (24) & 0.13 $\pm$ 0.17 (34) & ... & 0.05 $\pm$ 0.18 (168)\\
   & Cannon &-0.08 $\pm$ 0.20 (79) & -0.04 $\pm$ 0.21 (54) & -0.01 $\pm$ 0.21 (27) & 0.18 $\pm$ 0.31 (49) & ... & -0.01 $\pm$ 0.25 (209)\\
\hline
Co & DR13 &-0.06 $\pm$ 0.12 (78) & ... & ... & 0.09 $\pm$ 0.14 (104) & ... & 0.03 $\pm$ 0.15 (182)\\
   & DR14 &0.03 $\pm$ 0.14 (79) & ... & ... & 0.22 $\pm$ 0.27 (196) & ... & 0.15 $\pm$ 0.24 (275)\\
   & Cannon &0.01 $\pm$ 0.17 (79) & ... & ... & 0.19 $\pm$ 0.47 (222) & ... & 0.15 $\pm$ 0.42 (301)\\
\hline
\enddata    
\end{deluxetable*}

\subsection{Carbon, C}\label{sec:c}
In ASPCAP, the carbon abundance is determined in three ways: firstly, [C/M] is determined as one of the stellar parameters from fitting of the entire spectra. Secondly, once the stellar parameters have been fixed, the carbon abundance is determined from wide windows of the spectra covering numerous CN and CO molecular lines, among these the $^{13}$CO 3-0 molecular band head and several $^{12}$CO band heads (4-1, 5-2, 6-3, 7-4, 8-5, 9-6), that are sensitive to the derived effective temperature as well as surface gravity. Thirdly, the carbon abundance is derived using mainly six regions with C I lines (the lines at 15784.5~\AA, 16004.9~\AA, 16021.7~\AA, 16333.9~\AA, 16505.2~\AA, and 16890.4~\AA). The last two methods are expected to give more accurate carbon abundances, and calibrated values are only given for those two methods. Therefore, those are used in the comparison below. However, the differences between the carbon abundances derived from the initial `parameter' run based on the entire spectra, and the subsequent run using solely features from carbon-bearing molecules, are only a few hundredths of a dex for the stars considered here.

\begin{figure*}
\epsscale{1.1}
\plotone{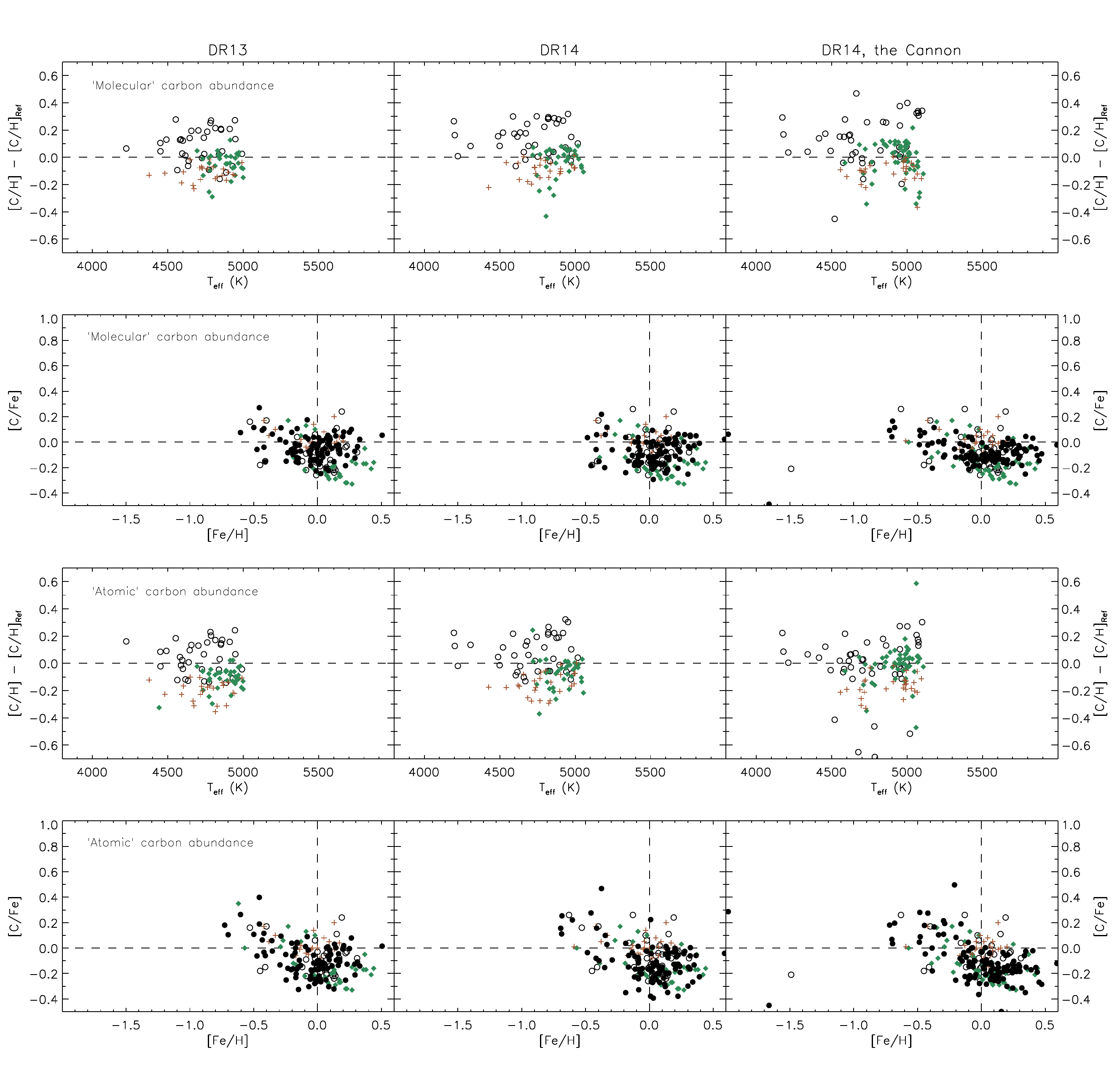}
\caption{The first and third rows shows differences in carbon abundance for the same stars in different analyses, and the second and fourth rows shows [C/Fe] vs. [Fe/H] for the same stars in different analyses. In the upper two rows carbon abundances as derived from CN and CO molecules in ASPCAP are showed, while the lower two rows show the same thing, but with ASPCAP carbon abundances derived from atomic C I lines instead. The \citet{2016ApJS..225...32B}-stars are marked using green diamonds, the \citet{2015A&A...580A..24D}-stars are marked using brown crosses, the values from Gaia-ESO DR3 are marked using black open circles, and the APOGEE results are marked using black filled circles.\label{fig:c}}
\end{figure*}

When comparing the references with carbon abundances determined to the molecular carbon abundances of APOGEE (see the top row of panels in Figure \ref{fig:c}), the DR13 [C/H] are systematically 0.02 dex lower than the references and the abundance difference show a spread  of 0.13 dex. In DR14, the systematic shift is 0.01 dex and the spread 0.15 dex. The carbon abundances from the Cannon, show a surprisingly large spread of 0.29 dex.

The third row of panels in Figure \ref{fig:c} shows the same comparison, but using the carbon abundance as derived from the C I lines instead. The results are similar, with DR13 showing a systematic shift of -0.10 dex and a spread of 0.14 dex, DR14 shows a shift of -0.06 dex and a spread of 0.15 dex, and the Cannon shows a shift of -0.03 dex and a spread of 0.22 dex.

However, as is obvious from the first and third row of panels in Figure \ref{fig:c}, there seem to be systematic differences among the high resolution optical studies, with, for example, the carbon abundances of Gaia-ESO being systematically lower compared to ASPCAP, and the carbon abundances of \citet{2015A&A...580A..24D} being systematically higher compared to ASPCAP.  With the seemingly large systematic differences between the comparison samples, it is far from obvious which of the two ASPCAP-derived carbon abundances is most accurate; the one derived from molecular lines or that from atomic lines. Nevertheless, looking at the individual references in Table \ref{tab:abunds}, gives the impression that the molecular carbon abundance is closer to the reference-values in two of the three cases \citep{2016ApJS..225...32B,2015A&A...580A..24D} which might indicate that the molecular carbon abundance reported by APOGEE is more accurate than the atomic carbon abundance. This is possibly corroborated by comparing the APOGEE molecular carbon trend with the atomic carbon trend in rows two and four of Figure \ref{fig:c}, respectively, where the molecular trends are tighter, especially for the Cannon analysis. However, since the stars for which abundances are shown in this plot are all differently evolved giants, and hence are expected to have different amount of CN-processed material in their photospheres, it is far from certain that the trend is \emph{expected} to be tight. Furthermore, we note that the number of lines available for `molecular' determination of carbon is much larger than for `atomic' carbon determination, and hence the abundance from molecular features is at least expected to be better for low S/N spectra. 

Regarding the cosmic origin of carbon, there are two carbon isotopes of astrophysical interest with different origin: $^{12}$C and $^{13}$C. $^{12}$C is formed via the triple-alpha process in helium burning and, on a cosmic scale, about half of the $^{12}$C is formed in massive stars and released into the interstellar medium (ISM) by type II supernovae (SNeII), and about half the $^{12}$C is formed in low-mass asymptotic giant branch (AGB) stars and is released into the ISM by stellar winds. $^{13}$C is formed in the CN-cycle, and mainly by intermediate-mass AGB stars. The $^{12}$C/$^{13}$C ratio in the photosphere of a star is expected to increase as the star ascends the giant branch, since material from deeper, hotter layers, where $^{12}$C can be turned into $^{13}$C via proton capture, are dredged up \citep{2003hic..book.....C,2014PASA...31...30K}. In future data releases, we hope to provide $^{12}$C/$^{13}$C ratios.
 
\subsection{Nitrogen, N}
The nitrogen abundance is determined two times within ASPCAP, firstly from the entire spectra as one of the stellar parameters, and then once more during the abundance determinations, from wide regions of the spectra covering numerous CN molecular lines. Many of these lines are sensitive to the derived surface gravity.  As with carbon, we use the second `non-parameter' nitrogen determination in the comparisons below. However, just like for carbon, the differences for the two nitrogen abundances are only of the order of a few hundredths of a dex for the stars considered here.

\begin{figure*}
\epsscale{1.1}
\plotone{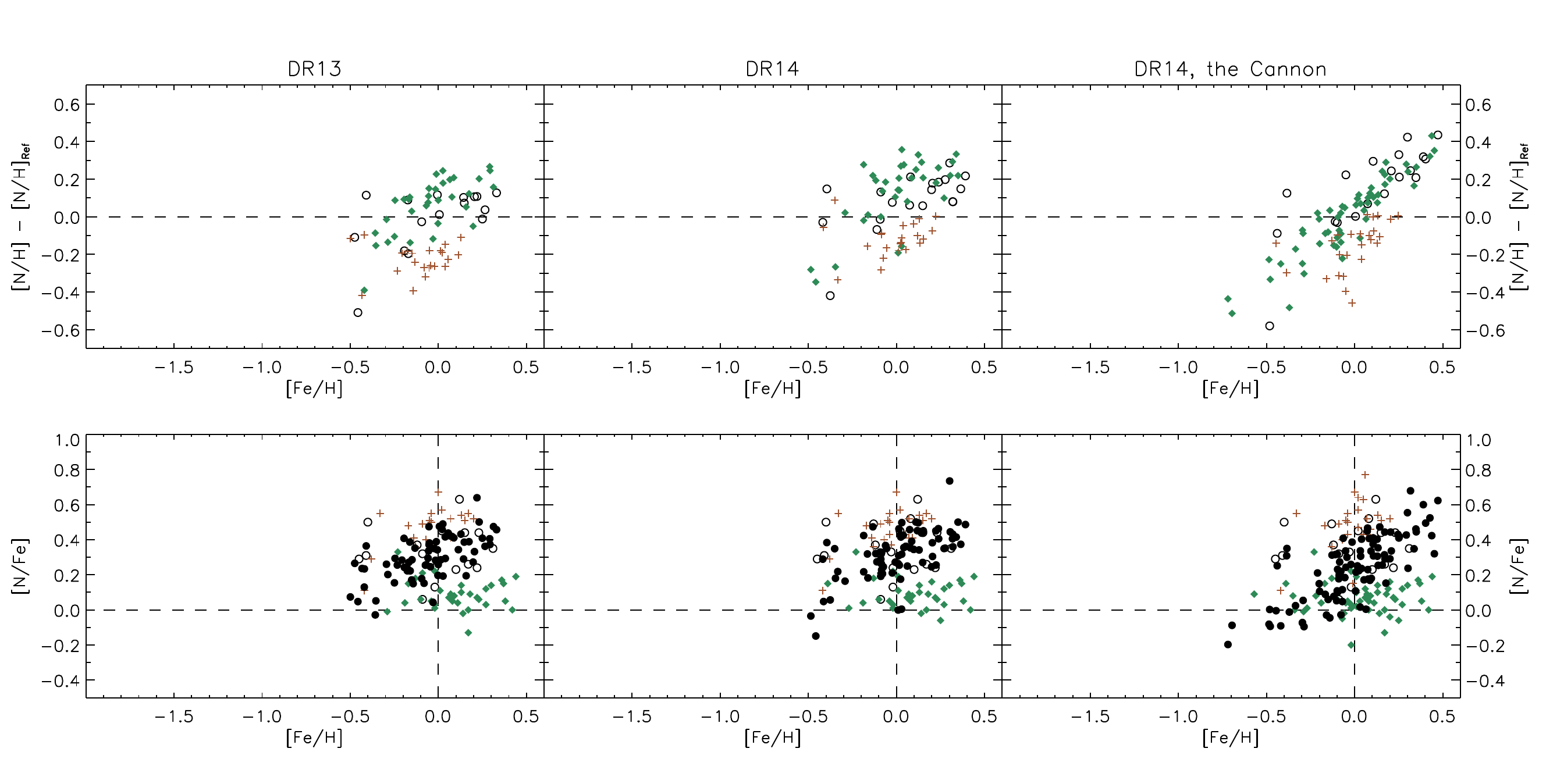}
\caption{The first row shows differences in nitrogen abundance for the same stars in different analyses, and the second row shows [N/Fe] vs. [Fe/H] for the same stars in different analyses. The \citet{2016ApJS..225...32B}-stars are marked using green diamonds, the \citet{2015A&A...580A..24D}-stars are marked using brown crosses, the values from Gaia-ESO DR3 are marked using black open circles, and the APOGEE results are marked using black filled circles.\label{fig:n}}
\end{figure*}

When comparing the references with nitrogen abundances determined to the nitrogen abundances of APOGEE (see the top row of panels in Figure \ref{fig:n}), a very clear trend with [Fe/H] can be seen in all analyses, especially in the Cannon analysis. This trend could be a consequence of the trend of effective temperature with metallicity in DR13 and DR14. Since CN molecular lines influence much of the APOGEE spectra, the reason for this trend has to be tracked down: is it due to systematics in the references or in ASPCAP? We will attend to this issue in coming works.

In the bottom row of panels in Figure \ref{fig:n}, a seemingly systematic difference between the nitrogen abundances derived in \citet{2016ApJS..225...32B} and \citet{2015A&A...580A..24D} can be seen. However, since the \citet{2015A&A...580A..24D} stars are higher on the giant branch than the \citet{2016ApJS..225...32B}-stars, this difference might be the sign of convective motions dredging up nitrogen into the atmosphere of the more evolved giants.

Nitrogen is produced in the CN-cycle, and on a cosmic scale, by intermediate-mass AGB stars and is released into the ISM via their stellar winds \citep{2003hic..book.....C,2014PASA...31...30K}.

\subsection{Oxygen, O}\label{sec:o}
In ASPCAP, the oxygen abundance is determined from 70 regions of the spectra covering numerous OH molecular lines. The OH-lines, and thus the oxygen abundance derived from them, are very sensitive to the determined effective temperature.

\citet{2015A&A...576A.128D} examine the 3D/NLTE-effects in H-band OH-lines for extremely metal poor stars ([Fe/H]$\sim-3$), and find negative corrections of about $-0.2$ dex. However, they do not specify any expected corrections for more metal-rich stars, like the bulk of the APOGEE sample. \citet{2004A&A...417..751A} investigate and compare different oxygen diagnostics  -- optical O I lines, optical [O I] lines, OH vib-rot lines around 3 $\mu$m, OH rot-rot lines around 9-13 $\mu$m -- and need to use 3D/NLTE modeling to make them agree, something that might influence both the APOGEE and reference analyses, all performed in 1D LTE.

\begin{figure*}
\epsscale{1.1}
\plotone{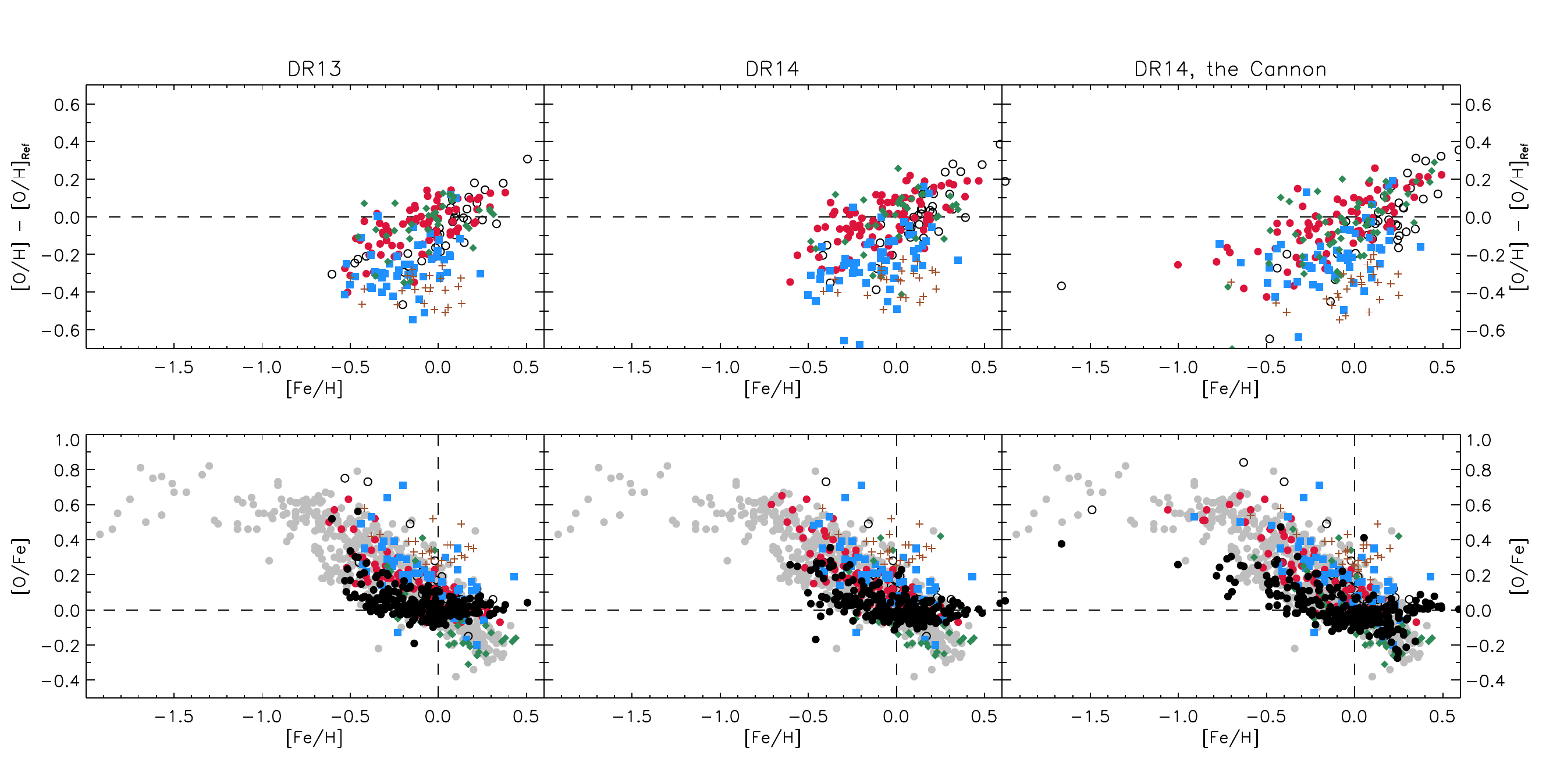}
\caption{The first row shows differences in oxygen abundance for the same stars in different analyses, and the second row shows [O/Fe] vs. [Fe/H] for the same stars in different analyses. The BACCHUS analyzed ARCES-stars are marked using blue squares, the \citet{2016ApJS..225...32B}-stars are marked using green diamonds, the \citet{2015A&A...580A..24D}-stars are marked using brown crosses, the values from Gaia-ESO DR3 are marked using black open circles, the \citet{2017A&A...598A.100J}-stars are marked using red dots, and the APOGEE results are marked using black filled circles. In the bottom row panels, the values from \citet{2014A&A...562A..71B} are shown in the background using gray dots.\label{fig:o}}
\end{figure*}

Oxygen is produced by massive stars through helium burning, and on a cosmic scale, it is released to the ISM via SNeII \citep{2003hic..book.....C}, and is expected to show an alpha-typical `knee'-like behavior in a [O/Fe] vs. [Fe/H] plot. There is a strong trend between the ASPCAP-derived oxygen abundance and the references with [Fe/H] in the first row of panels in Figure \ref{fig:o}. Just like for nitrogen, this trend might in fact be due to the temperature trend with metallicity affecting the determined oxygen abundance.

The bottom row of panels in Figure \ref{fig:o} shows the [O/Fe] vs. [Fe/H] trends for the various analyses, and the trend of determined oxygen abundance with [Fe/H] from the top row of panels is reflected in that the APOGEE [O/Fe] vs. [Fe/H] trends do not reach as high [O/Fe] for low metallicites as the reference works. 

\subsection{Sodium, Na}
In ASPCAP, the sodium abundance is determined from two weak (in GK-giants) and possibly blended lines at 16373.9~\AA~and 16388.9~\AA.

\citet{2016ApJ...830...35S} use the same lines, and derive sodium abundances about 0.2 dex higher than the calibrated DR13 abundances in their manual re-analysis of DR13 APOGEE spectra of 12 giants in NGC~2420 ([Fe/H]$\sim -0.16$). \citet{2015ApJ...798L..41C} discuss the NLTE effects for these lines for the 11 metal-rich giants from NGC 6791 ([Fe/H]$\sim +0.3$) and find them to be very small (maximum 0.04 dex). \citet{2011A&A...528A.103L}, however, show that NLTE corrections might be large for certain combinations of stellar parameters for several of the often used optical spectral lines, something that possibly might influence the accuracy of some of the comparison works.

\begin{figure*}
\epsscale{1.1}
\plotone{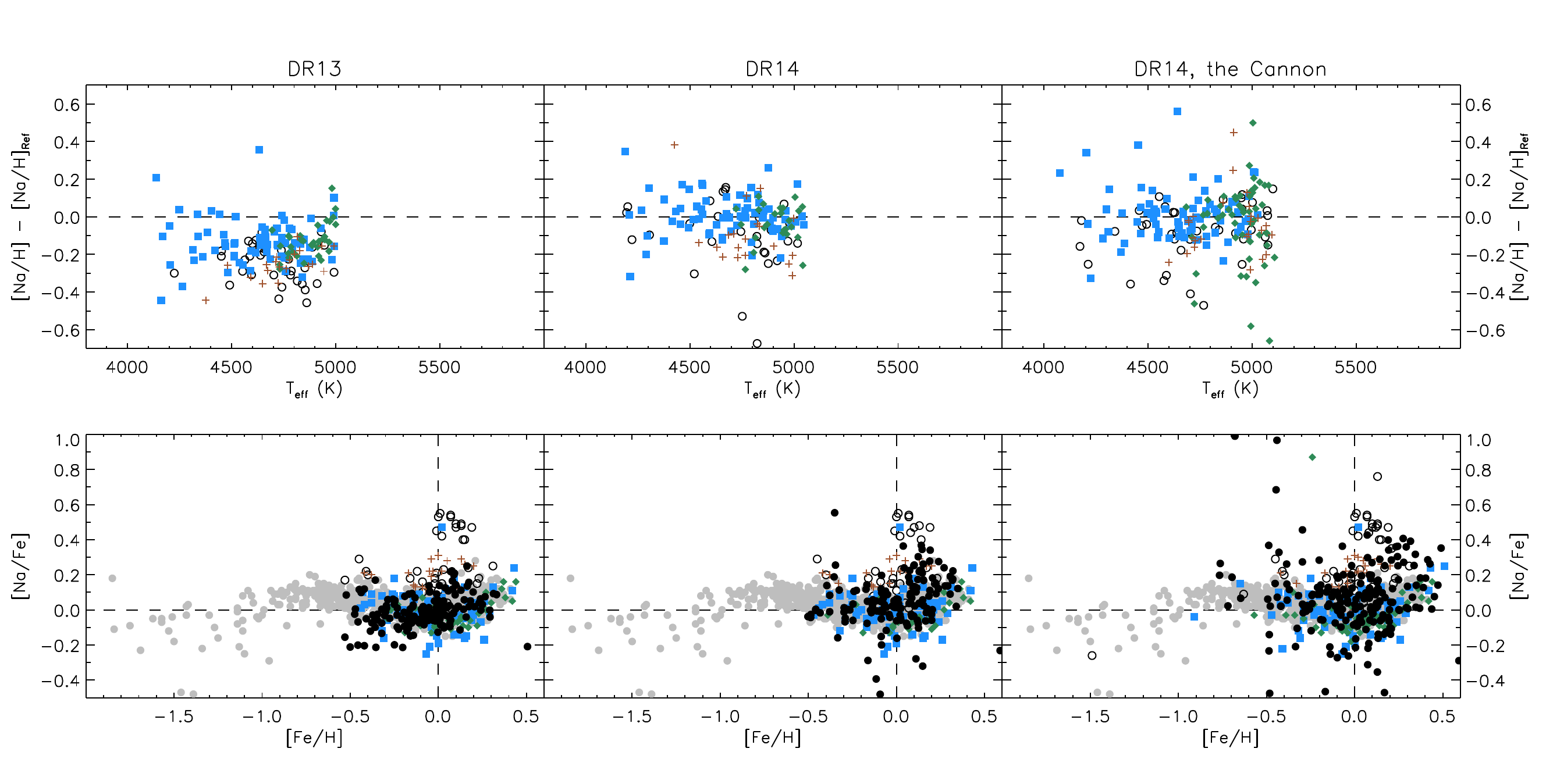}
\caption{The first row shows differences in sodium abundance for the same stars in different analyses, and the second row shows [Na/Fe] vs. [Fe/H] for the same stars in different analyses. The BACCHUS analyzed ARCES-stars are marked using blue squares, the \citet{2016ApJS..225...32B}-stars are marked using green diamonds, the \citet{2015A&A...580A..24D}-stars are marked using brown crosses, the values from Gaia-ESO DR3 are marked using black open circles, and the APOGEE results are marked using black filled circles. In the bottom row panels, the values from \citet{2014A&A...562A..71B} are shown in the background using gray dots.\label{fig:na}}
\end{figure*}

Compared to the references (see the top row of panels in Figure \ref{fig:na}), the DR13 [Na/H] are systematically 0.18 dex lower and the abundance differences show a spread of 0.19 dex. In DR14, the systematic shift is -0.03 dex and the spread 0.16 dex. The sodium abundances from the Cannon show a systematic shift of -0.02 dex and a significantly larger spread of 0.24 dex as compared to the references. The \citet{2015A&A...580A..24D}-stars in general have the highest sodium abundances, see the second row of panels in Figure \ref{fig:na}. In general, the [Na/Fe] vs. [Fe/H] trends for the giants follow the trend of the dwarf stars of \citet{2014A&A...562A..71B}.

Sodium is mainly produced by explosive carbon burning in SNeII, and is in this process deposited into the ISM. However, about a tenth of the sodium in the cosmos is instead produced in helium burning shells of evolved lower-mass stars (especially in the more massive AGB stars), and later deposited into the ISM through stellar winds \citep{2003hic..book.....C,2014PASA...31...30K}. 

\subsection{Magnesium, Mg}
In ASPCAP, the magnesium abundance is determined from 14 Mg I-lines of different strengths, some of which are blended, and some which seem mostly unblended (in GK-giants). Three of the lines are strong in giants and have pressure-broadened wings, meaning that the determined magnesium abundance is likely to be sensitive to the derived surface gravity.

\begin{figure*}
\epsscale{1.1}
\plotone{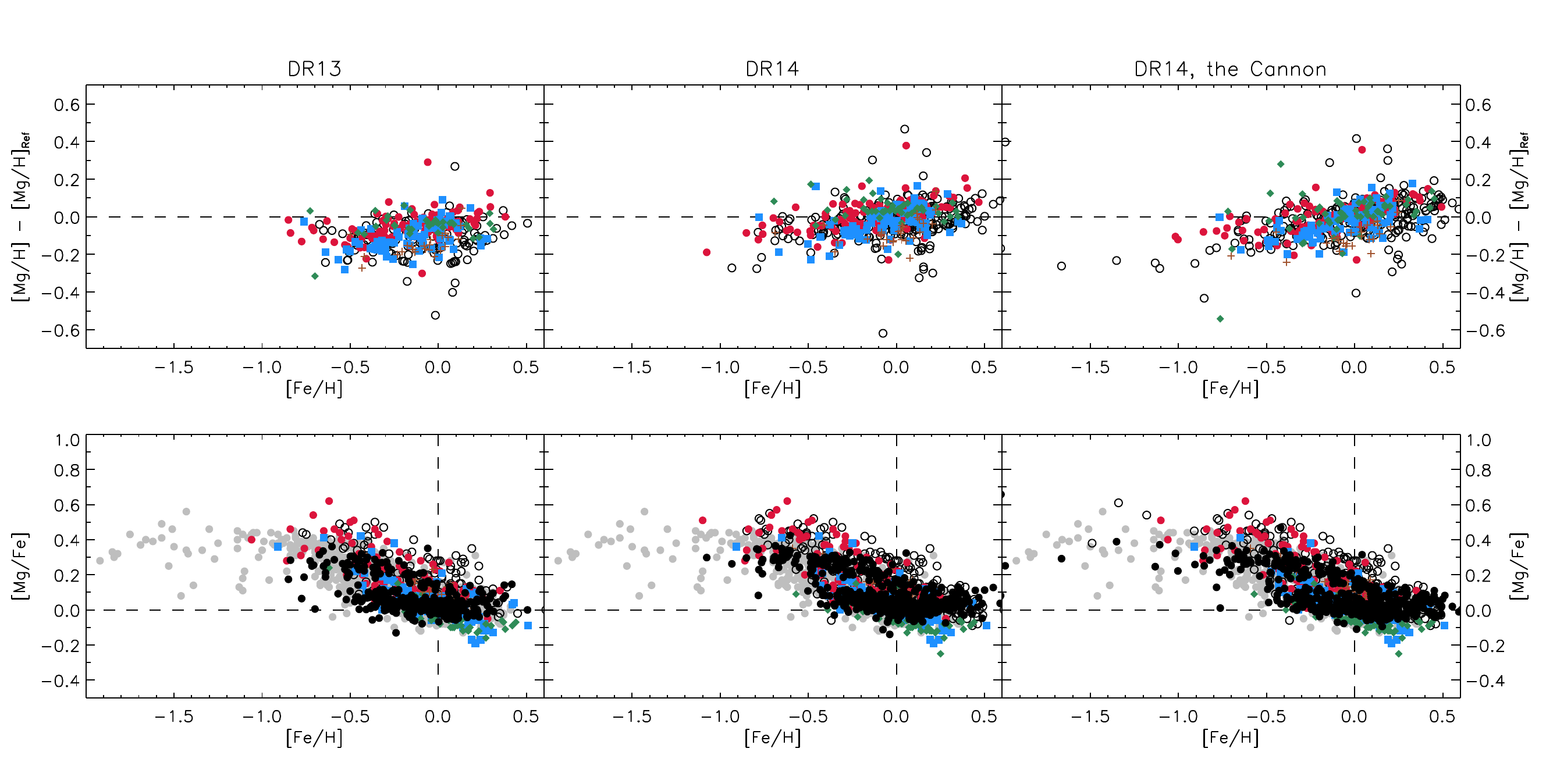}
\caption{The first row shows differences in magnesium abundance for the same stars in different analyses, and the second row shows [Mg/Fe] vs. [Fe/H] for the same stars in different analyses. The BACCHUS analyzed ARCES-stars are marked using blue squares, the \citet{2016ApJS..225...32B}-stars are marked using green diamonds, the \citet{2015A&A...580A..24D}-stars are marked using brown crosses, the values from Gaia-ESO DR3 are marked using black open circles, the \citet{2017A&A...598A.100J}-stars are marked using red dots, and the APOGEE results are marked using black filled circles. In the bottom row panels, the values from \citet{2014A&A...562A..71B} are shown in the background using gray dots.\label{fig:mg}}
\end{figure*}

Compared to the reference studies, the DR13 magnesium abundances are 0.09 dex lower with a spread of 0.10 dex, while for DR14, the systematic shift is -0.01 dex and the spread is 0.09 dex, making magnesium the alpha-element most accurately determined by APOGEE (as compared to the references). There might however be a hint of a weak trend with [Fe/H], see the first row of panels in Figure \ref{fig:mg}.

\citet{2017ApJ...835...90Z} evaluated the NLTE-effects for eight H-band Mg I lines, finding relatively large negative corrections for the three strong lines at 15740.7~\AA, 15749.0~\AA, and 15765.8~\AA~of about -0.15 dex for GK-giants, and larger corrections for stars higher up the giant branch. If applicable to all of the H-band lines used in the ASPCAP analysis, this negative NLTE-correction would mean that the magnesium abundance derived by APOGEE would be overestimated, but the opposite is suggested from the comparison with the references, especially for the metal-poor stars.

Magnesium is produced via carbon burning, and on a cosmic scale, magnesium is an alpha-element mainly returned to the ISM through SNeII \citep{2003hic..book.....C}. As such it is expected to show the typical `knee'-like behavior in a [Mg/Fe] vs. [Fe/H] plot, which it does in all analyses, see the bottom row of panels of Figure \ref{fig:mg}. Also, the distinction between the thin and the thick disk abundance patterns is obviously visible in all analyses.

\subsection{Aluminium, Al}
In ASPCAP, the aluminium abundance is determined mainly from three regions of the spectra covering the Al I lines around 16718.9~\AA, 16750.5~\AA, and 16763.4~\AA. The lines are sensitive to the derived effective temperature, and since they are rather strong, they are also sensitive to the adopted surface gravity and microturbulence.

In \citet{2013ApJ...765...16S,2016ApJ...830...35S}, only the two Al I lines at 16718.9~\AA~and 16763.4~\AA~are used. \citet{2016ApJ...830...35S} derived aluminum abundances about 0.15 dex higher compared to calibrated DR13 abundances in their manual re-analysis of APOGEE spectra of 12 giants in NGC~2420 ([Fe/H]$\sim -0.16$). 

\citet{2016A&A...594A..43H} use only one line at 16763.4~\AA, finding that the 16718.9~\AA~line is poorly fit in the core, and suggest this is due to NLTE effects. This suspicion is corroborated by \citet{2017A&A...607A..75N}, showing that the NLTE effects for this line can be of the order of 0.2 dex, depending on the stellar parameters. Possibly for this reason, \citet{2016A&A...594A..43H} find about 0.1 dex lower aluminium abundances for metal-poor stars compared to DR12.

\begin{figure*}
\epsscale{1.1}
\plotone{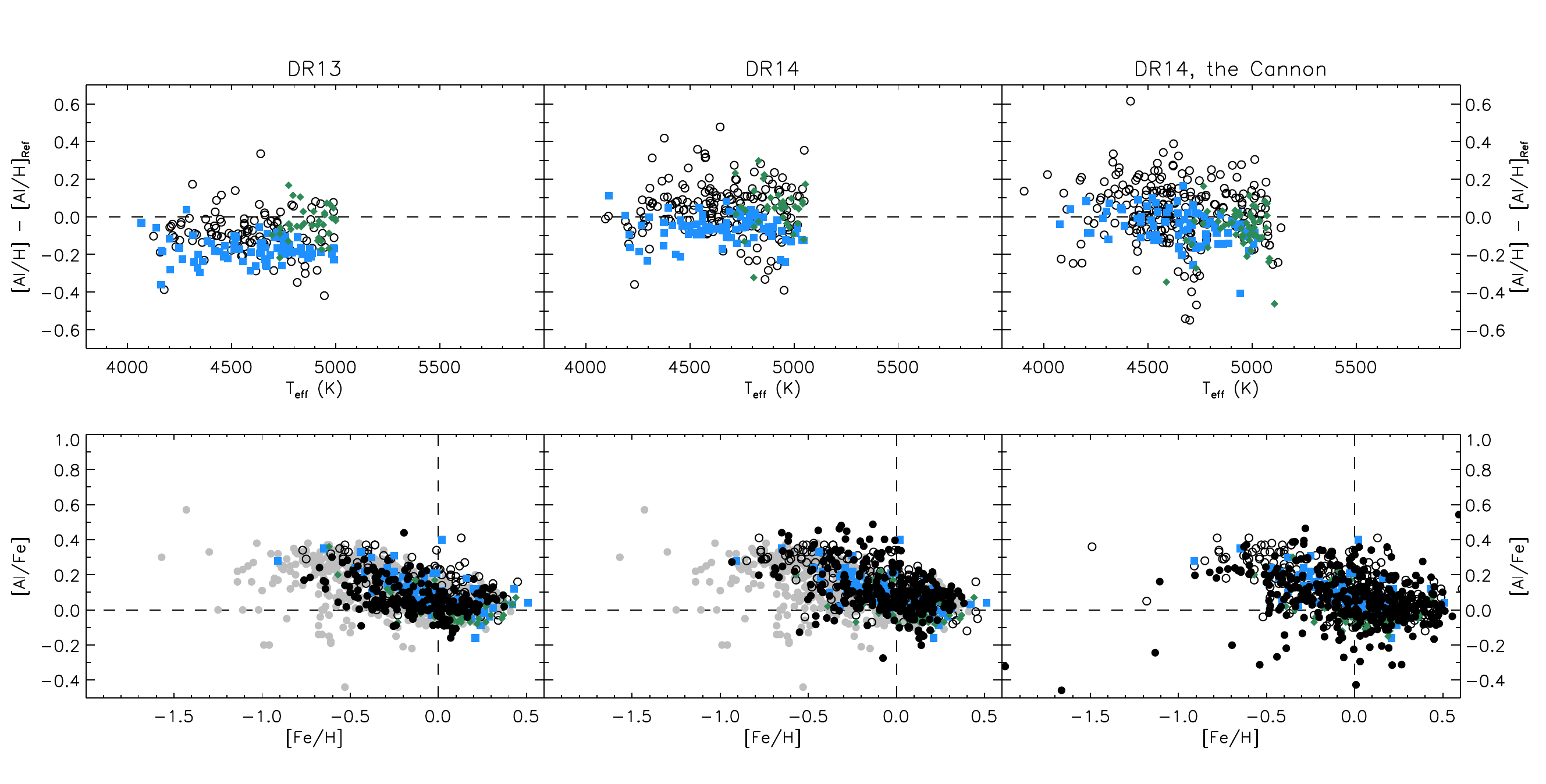}
\caption{The first row shows differences in aluminium abundance for the same stars in different analyses, and the second row shows [Al/Fe] vs. [Fe/H] for the same stars in different analyses. The BACCHUS analyzed ARCES-stars are marked using blue squares, and the \citet{2016ApJS..225...32B}-stars are marked using green diamonds, the values from Gaia-ESO DR3 are marked using black open circles, and the APOGEE results are marked using black filled circles. In the bottom row panels, the values from \citet{2014A&A...562A..71B} are shown in the background using gray dots.\label{fig:al}}
\end{figure*}

The DR13 aluminium abundances are 0.12 dex lower than the references, with a spread of 0.15 dex. For DR14, the systematic shift is 0.01 dex and the spread is 0.14 dex, as shown in the first row of panels in Figure \ref{fig:al}.

Aluminium is formed by carbon burning in massive stars, and on a cosmic scale it is released into the ISM by SNeII \citep{2003hic..book.....C}. Hence, as one would expect, the [Al/Fe] vs. [Fe/H]-trend shows an alpha-like behavior in \citet{2014A&A...562A..71B} and also in all APOGEE analyses in the second row of panels in Figure \ref{fig:al}.

\subsection{Silicon, Si}
In ASPCAP, the silicon abundance is determined from 17 Si I-lines of different strengths, some of which are blended, and some of which seem mostly unblended (in GK-giants). 

\citet{2016A&A...594A..43H} find systematically 0.2 dex lower silicon abundances compared to DR12, when using only the five lines at 15376.8~\AA, 15888.4~\AA, 16215.7~\AA, 16680.8~\AA, and 16828.2~\AA. 

\begin{figure*}
\epsscale{1.1}
\plotone{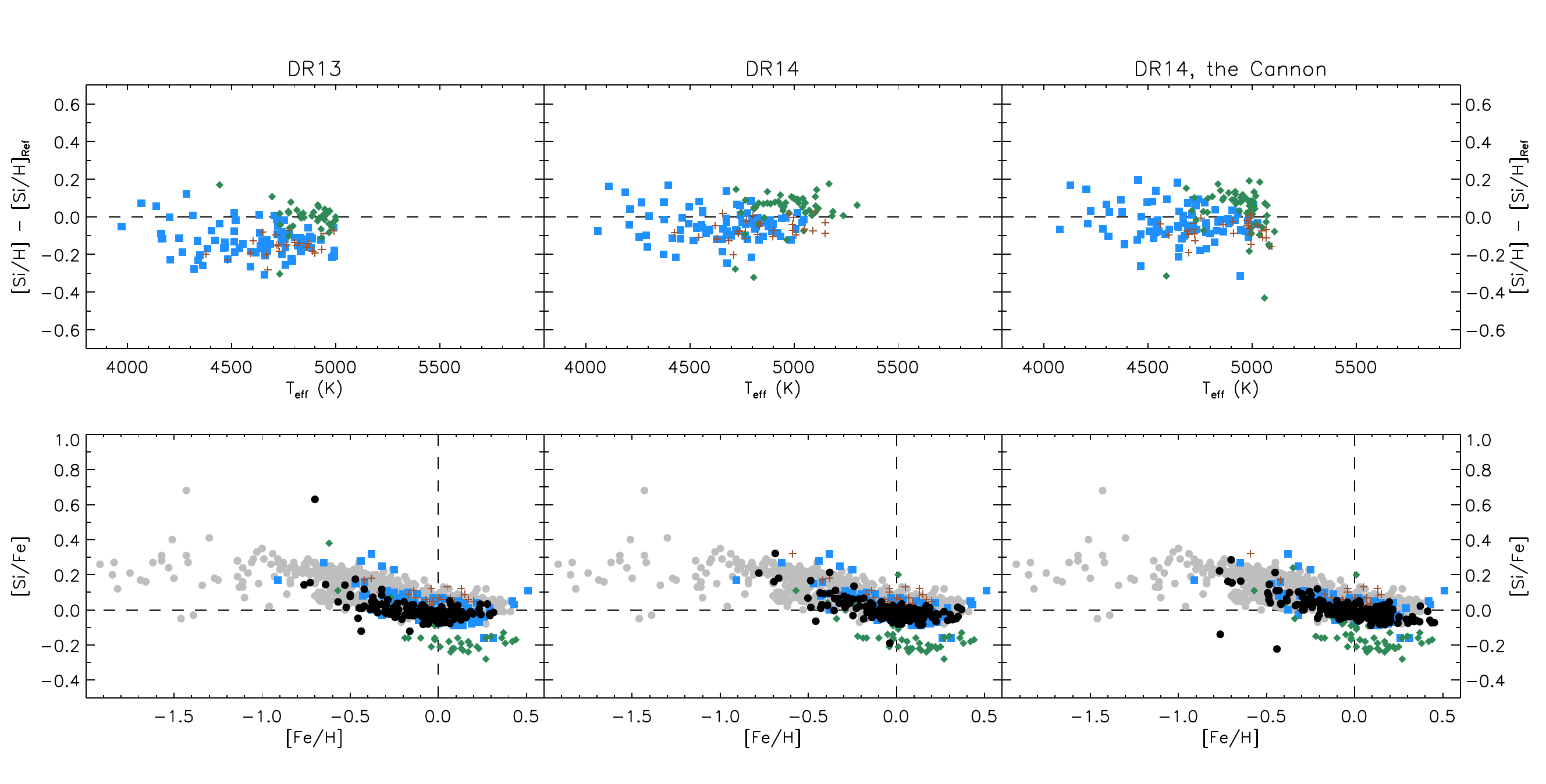}
\caption{The first row shows differences in silicon abundance for the same stars in different analyses, and the second row shows [Si/Fe] vs. [Fe/H] for the same stars in different analyses. The BACCHUS analyzed ARCES-stars are marked using blue squares, and the \citet{2016ApJS..225...32B}-stars are marked using green diamonds, the \citet{2015A&A...580A..24D}-stars are marked using brown crosses, and the APOGEE results are marked using black filled circles. In the bottom row panels, the values from \citet{2014A&A...562A..71B} are shown in the background using gray dots.\label{fig:si}}
\end{figure*}

In a global comparison with the references, the DR13 silicon abundances are 0.12 dex lower, with a spread of 0.10 dex, while for DR14, the systematic shift is -0.03 dex and the spread is 0.11 dex (see the first row of panels in Figure \ref{fig:si}).
 
\citet{2016ApJ...833..137Z} predict that NLTE-corrections for the H-band lines in GK-giants should be of the order of -0.2 dex for the two strong lines at 15888.4~\AA~and 16680.8~\AA, but smaller for the other two lines investigated at 16380.2~\AA~and 16828.2~\AA. However, they are also negative, meaning that the silicon abundance derived from those lines assuming LTE would be overestimated, at odds with what we are finding when comparing to the references. \citet{2016ApJ...833..137Z} also found that NLTE-corrections should increase with decreasing surface gravities, making the situation more severe for the APOGEE targets on the top of the giant branch.

Silicon is mainly produced in oxygen burning, and on a cosmic scale it is deposited into the ISM by both SNeII and supernovae type Ia (SNeIa) \citep{2003hic..book.....C}. It is expected to show an alpha-typical `knee'-like trend in a [Si/Fe] vs. [Fe/H] plot. This shape is visible in all references and APOGEE analyses, but there are some systematic uncertainties affecting the silicon determination in the giant stars of \citet{2016ApJS..225...32B}, shifting the trend of the green diamonds downwards (second row of panels in Figure \ref{fig:si}). The trends in all APOGEE analyses are tight, but do not clearly show the separation of the thin and thick disk type abundances. Surprisingly, there are two outliers in the otherwise very tight Cannon-trend, which are not present in the DR14 analysis.

\subsection{Phosphorus, P}
In ASPCAP, the phosphorus abundance is determined from three P I lines that are all blended: 15711.5~\AA, blended with a Fe I line, 16254.7~\AA, blended with an OH molecular line, and 16482.9~\AA, blended with a CO molecular line.

\citet{2016A&A...594A..43H} use only the two lines at 15711.5~\AA~and 16482.9~\AA~to derive upper limits for the phosphorus abundances.

Phosphorus abundances have not been determined in any of the comparison works, and not much work has been done on the galactic chemical evolution of phosphorus, mainly because there are no optical spectral lines. On a cosmic scale, phosphorus is believed to be formed by carbon and neon burning in massive stars \citep{2003hic..book.....C}, but presently galactic chemical evolution models have difficulties to fit the observations \citep{2017ApJ...841..108M}, leaving the actual origin of phosphorus relatively uncertain. 

\subsection{Sulfur, S}
In ASPCAP, the sulfur abundance is determined from three S I lines, but for two of the lines, the number of usable data points in the reduced spectra -- pixels in the APOGEE apStar files -- are very low: from the line at 15403.8~\AA, one pixel is used, and from the line at 16576.6~\AA, two pixels are used. The last line at 15478.5~\AA~is blended with Fe I.

\citet{2016A&A...594A..43H} derive lower sulfur abundances compared to DR12, especially for the most metal-rich stars, where they derive sulfur abundances up to about 0.2 dex lower. They dismiss a non-ASPCAP S I line at 15469.8 (blended with OH) on account of suspected NLTE/3D effects, and only use the ASPCAP S I line at 15478.5~\AA~(blended with Fe I). 

\begin{figure*}
\epsscale{1.1}
\plotone{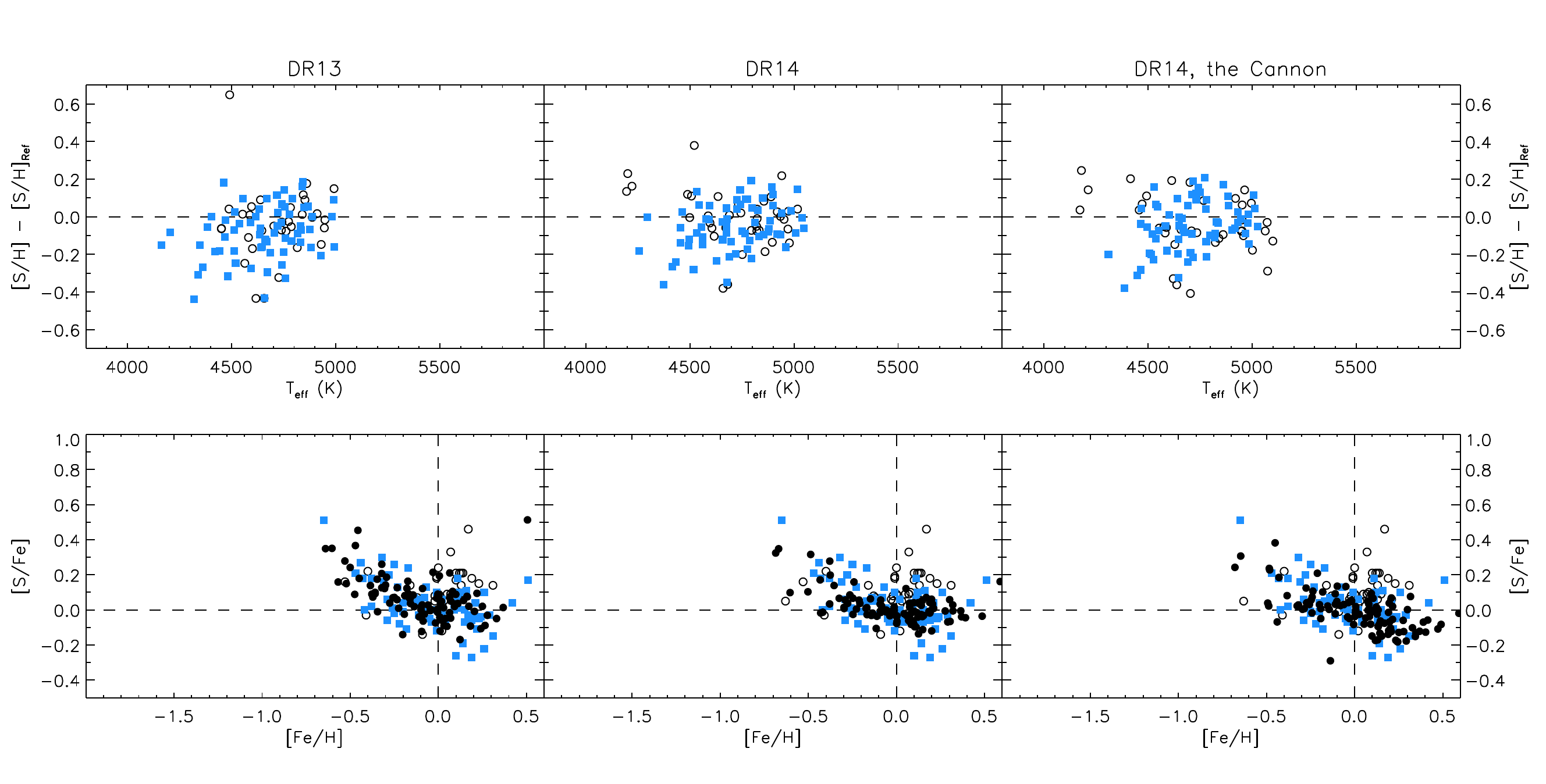}
\caption{The first row shows differences in sulfur abundance for the same stars in different analyses, and the second row shows [S/Fe] vs. [Fe/H] for the same stars in different analyses. The BACCHUS analyzed ARCES-stars are marked using blue squares, the values from Gaia-ESO DR3 are marked using black open circles, and the APOGEE results are marked using black filled circles.\label{fig:s}}
\end{figure*}

Comparing with the references, the DR13 sulfur abundances are 0.06 dex lower, with a spread of 0.16 dex. For DR14, the systematic shift is -0.03 dex and the spread is 0.13 dex, as illustrated in the first row of panels in Figure \ref{fig:s}.

Sulfur is produced via oxygen burning in massive stars, and then released into the ISM by SNeII \citep{2003hic..book.....C}, meaning that one would expect the alpha-typical `knee'-like trend in an [S/Fe] vs. [Fe/H] plot. However, this is not visible in the trends in the second row of Figure \ref{fig:s} since there are not metal-poor stars in the overlapping sample, instead just the decreasing part of the trend is shown. No obvious separation between thin and thick disk abundance patterns can be seen.

\subsection{Potassium, K}
In ASPCAP, the potassium abundance is determined from two K I lines with suitable strengths (in GK-giants) at 15163.1~\AA~and 15168.4~\AA~that are slightly blended with CN-lines.

On a cosmic scale, potassium is created in different amounts and ways in SNeII, depending on the mass of the progenitor, but is mainly the product of explosive oxygen burning \citep{2003hic..book.....C}.

\begin{figure*}
\epsscale{1.1}
\plotone{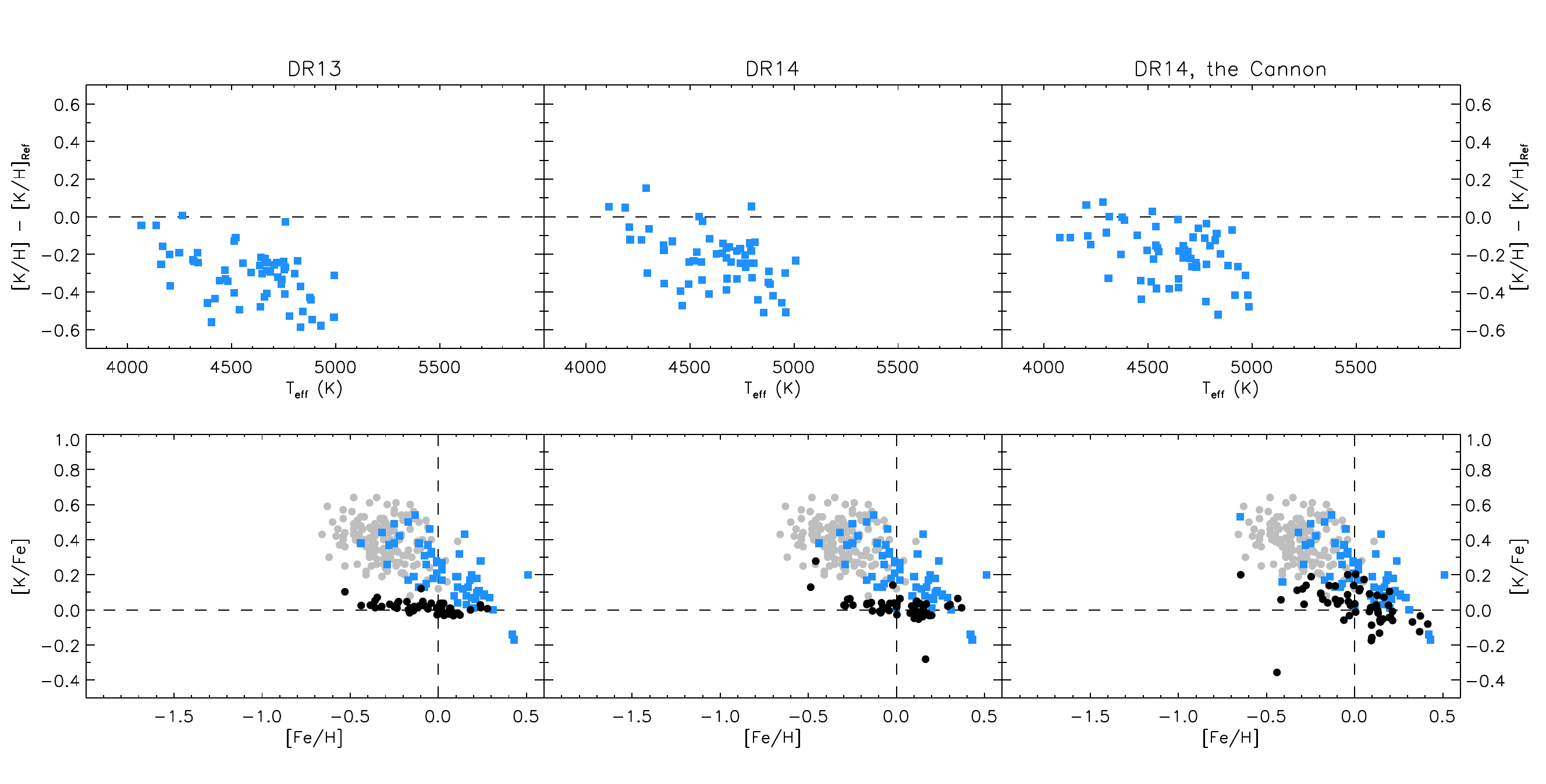}
\caption{The first row shows differences in potassium abundance for the same stars in different analyses, and the second row shows [K/Fe] vs. [Fe/H] for the same stars in different analyses. The BACCHUS analyzed ARCES-stars are marked using blue squares, and the APOGEE results are marked using black filled circles. In the bottom row panels, the values from \citet{2003MNRAS.340..304R} are shown in the background using gray dots.\label{fig:k}}
\end{figure*}

The potassium abundances from the APOGEE spectra and the reference show trends with both T$_{\mathrm{eff}}$ and metallicity, as is shown in the top row of panels in Figure \ref{fig:k}. The ASPCAP [K/Fe] vs. [Fe/H] trends are much tighter and show a different behavior than the reference, while the results from the Cannon show a larger spread and a trend more resembling the optical (see the second row of panels in Figure \ref{fig:k}). Since potassium is believed to be produced by SNeII, an alpha-like trend as seen in the BACCHUS analyzed ARCES-stars seem more probable than the ASPCAP-trends, even if the exact chemical evolution of potassium is rather unknown.

\subsection{Calcium, Ca}
In ASPCAP, the calcium abundance is determined from four Ca I lines at 16136.8~\AA, 16150.8~\AA, 16155.2~\AA, and 16157.4~\AA. The lines all are of suitable strengths and do not appear to be blended (in GK-giants).

\begin{figure*}
\epsscale{1.1}
\plotone{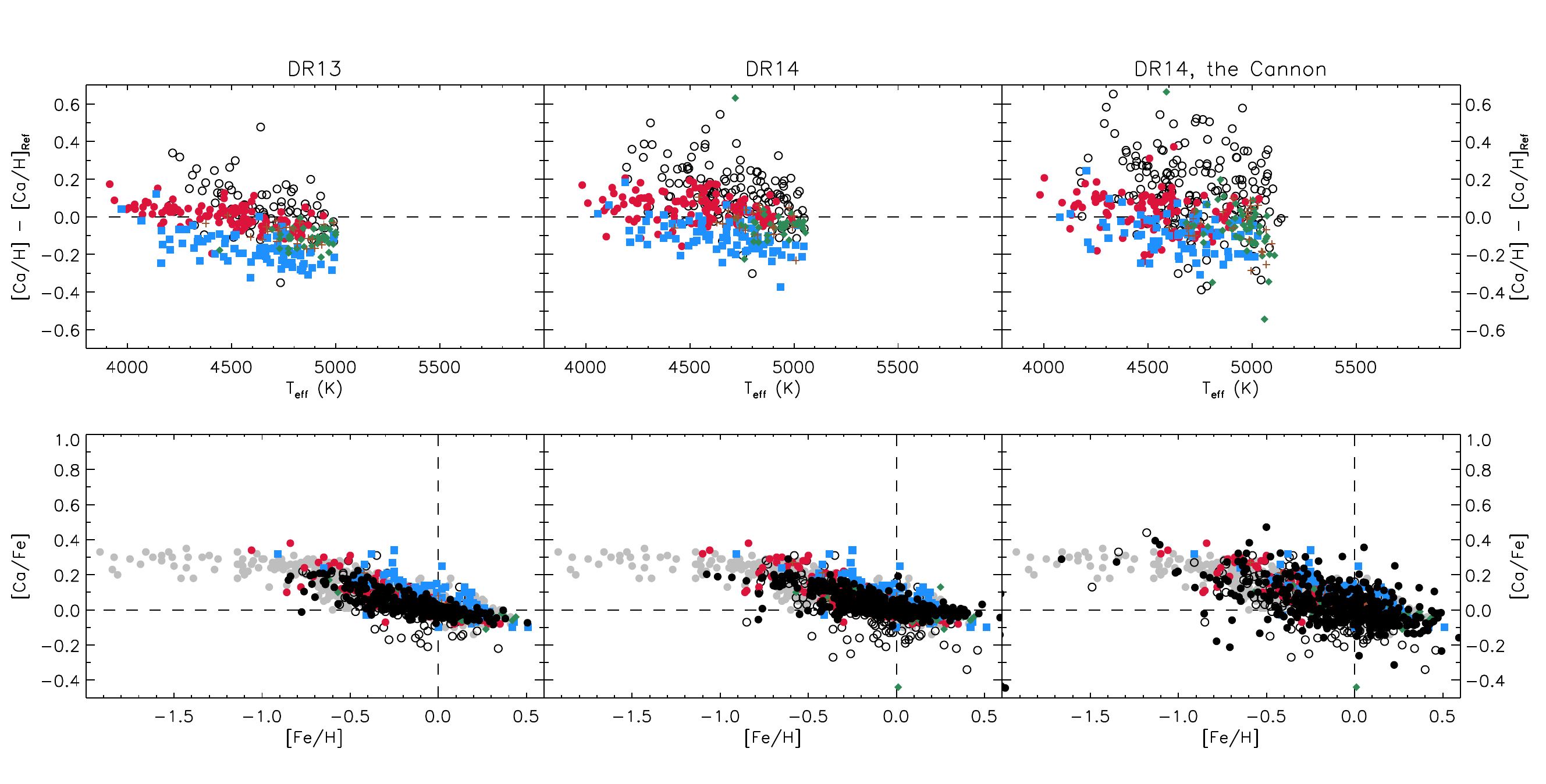}
\caption{The first row shows differences in calcium abundance for the same stars in different analyses, and the second row shows [Ca/Fe] vs. [Fe/H] for the same stars in different analyses. The BACCHUS analyzed ARCES-stars are marked using blue squares, the \citet{2016ApJS..225...32B}-stars are marked using green diamonds, the \citet{2015A&A...580A..24D}-stars are marked using brown crosses, the values from Gaia-ESO DR3 are marked using black open circles, the \citet{2017A&A...598A.100J}-stars are marked using red dots, and the APOGEE results are marked using black filled circles. In the bottom row panels, the values from \citet{2014A&A...562A..71B} are shown in the background using gray dots.\label{fig:ca}}
\end{figure*}

The DR13 calcium abundances are 0.04 dex lower than the references with a spread of 0.12 dex. For DR14, the systematic shift is +0.03 dex and the spread is 0.14 dex. There might also be a hint of a weak trend with T$_{\mathrm{eff}}$; see the first row of panels in Figure \ref{fig:ca}.

Calcium is produced in massive stars through oxygen burning and silicon burning, and on a cosmic scale, it is released into the ISM via SNeII \citep{2003hic..book.....C}. Calcium is therefore expected to show the alpha-typical `knee'-like behavior in a [Ca/Fe] vs. [Fe/H] plot, and it does in all analyses; see the bottom row of panels of Figure \ref{fig:ca}. However, the distinction between the thin and the thick disk abundance patterns is not nearly as obvious as for magnesium.

\subsection{Titanium, Ti}\label{sec:ti}
ASPCAP derives both Ti I and Ti II abundances.

All the nine Ti I-lines used except the one at 15315.6~\AA~are very sensitive to the adopted effective temperature. However, this line has a very low weight in the ASPCAP windows used for determining Ti I, and therefore it does not influence the determined Ti I abundance very much. Therefore, the derived Ti I-abundances in DR13/14 are expected to be very influenced by the trend of effective temperature with metallicity in the ASPCAP-analysis. Indeed, the Ti I abundance is the only abundance which -- according to our tests -- most certainly would benefit from a change in methodology in ASPCAP, to instead use calibrated stellar parameters and not, as currently done, use uncalibrated parameters when determining abundances.

The Ti I lines at 15334.8~\AA~and 15715.6~\AA~are excluded in \citet{2016A&A...594A..43H} due to showing a strange [Ti/Fe] vs. [Fe/H] trend (see their Fig. 2). Their proposed explanation for this is NLTE and/or saturation effects, but since \citet{2016A&A...594A..43H} use the DR10 effective temperatures in their analysis, these inaccurate trends are in fact likely due to the trend of effective temperature with metallicity in the ASPCAP-analysis.

The Ti II abundance is determined in APSCAP using a Ti II line at 15873.8~\AA~\citep{2014ApJ...787L..16W}, and this is the only titanium line not discarded by \citet{2016A&A...594A..43H} (see the first panel in their Fig. 2). This line is not sensitive to the adopted effective temperature, but instead to the surface gravity.

Titanium is formed by explosive silicon burning and fusion of helium-nuclei (to $^{48}$Cr that beta-decays to $^{48}$Ti). On a cosmic scale, titanium is mostly produced by SNeII \citep{2003hic..book.....C}, and as such it is expected to show the alpha-typical `knee'-like trend in a [Ti/Fe] vs. [Fe/H] plot.

\begin{figure*}
\epsscale{1.1}
\plotone{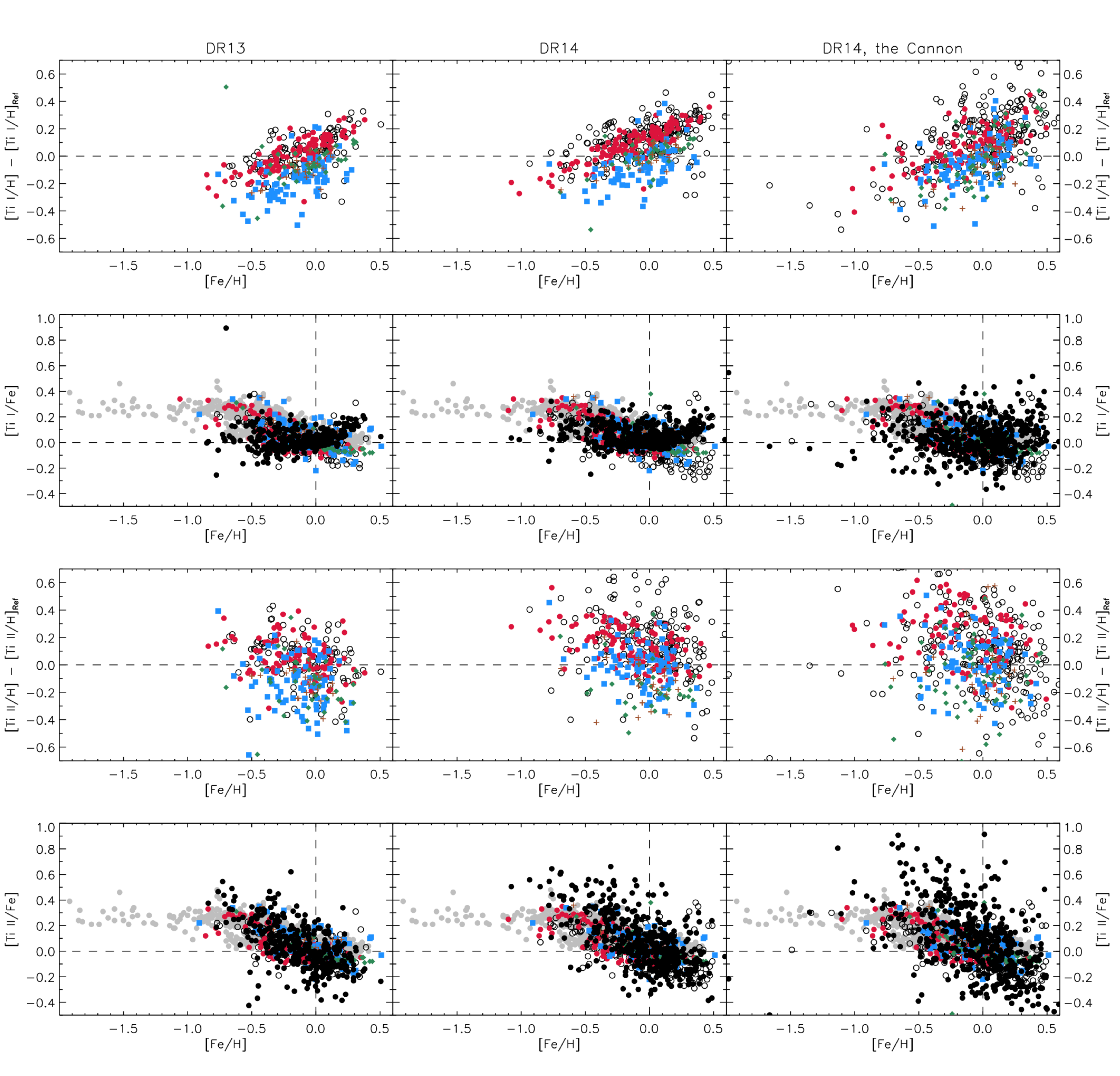}
\caption{The first row shows differences in Ti I abundance for the same stars in different analyses, and the second row shows [Ti I/Fe] vs. [Fe/H] for the same stars in different analyses. The third and fourth rows show the same, but for Ti II abundances. The BACCHUS analyzed ARCES-stars are marked using blue squares, the \citet{2016ApJS..225...32B}-stars are marked using green diamonds, the \citet{2015A&A...580A..24D}-stars are marked using brown crosses, the values from Gaia-ESO DR3 are marked using black open circles, the \citet{2017A&A...598A.100J}-stars are marked using red dots, and the APOGEE results are marked using black filled circles. In the bottom row panels, the values from \citet{2014A&A...562A..71B} are shown in the background using gray dots.\label{fig:ti}}
\end{figure*}

In the top row of panels in Figure \ref{fig:ti}, obvious trends of Ti I abundances vs. [Fe/H] can be seen. Since the derived titanium abundance is sensitive to the adopted T$_{\mathrm{eff}}$, this trend of titanium abundance difference with [Fe/H] could simply reflect the trend of effective temperature with [Fe/H]. In the second row of panels in Figure \ref{fig:ti}, the resulting inaccurate [Ti/Fe] vs. [Fe/H] trends are shown.

In the third row of panels in Figure \ref{fig:ti}, the Ti abundances from the Ti II line are shown. Since they are derived from a single line that happens to fall close to the gap between two of the detectors, continuum placement is challenging and the abundances are very uncertain, which is reflected by the very large scatter. However, in the bottom row of panels, the [Ti/Fe] vs. [Fe/H] plot  from the Ti II line shows the expected `knee' but, since the trend has a lot of dispersion, no clear separation of thin and thick disk abundance trends can be seen.

\subsection{Vanadium, V}
In ASPCAP, the vanadium abundance is determined from five (hyper fine splitted, hfs) V I lines at 15924.8~\AA, 16031.1~\AA~(heavily blended with CN/CO), 16200.2~\AA~(blended with CO), 16406.1~\AA~(heavily blended with CO), and 16570.6~\AA. 

\citet{2016A&A...594A..43H} and \citet{2016ApJ...830...35S} only use the 15924.8~\AA~line. \citet{2016A&A...594A..43H} derive higher vanadium abundances as compared to DR12 for the metal-poor stars (almost 0.2 dex for [Fe/H]$\sim-0.5$). \citet{2016ApJ...830...35S} derived vanadium abundances about 0.1 dex higher, as compared to calibrated DR13 abundances in their manual re-analysis of APOGEE spectra of 12 giants in NGC~2420 ([Fe/H]$\sim -0.16$). 

\begin{figure*}
\epsscale{1.1}
\plotone{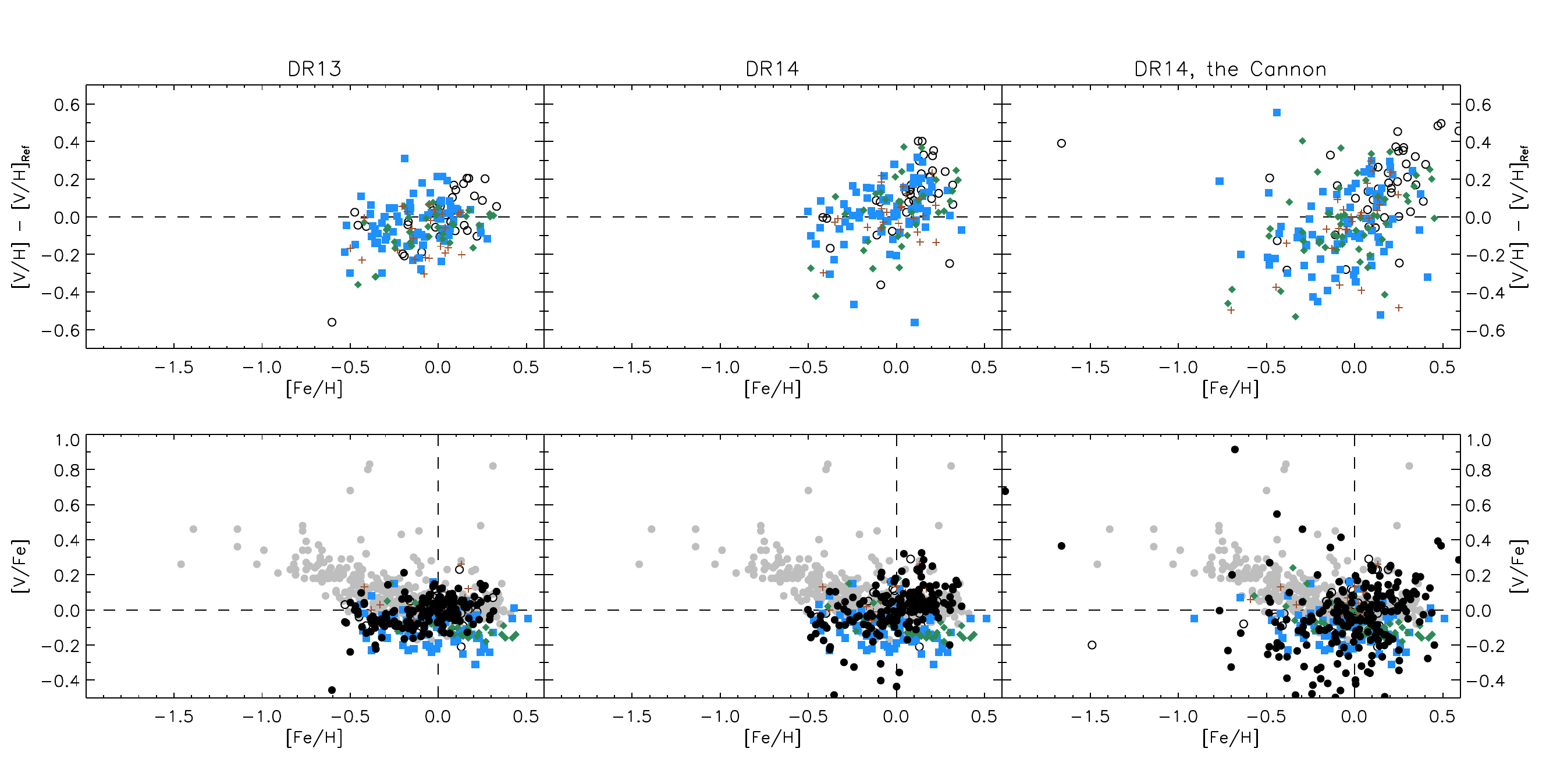}
\caption{The first row shows differences in vanadium abundance for the same stars in different analyses, and the second row shows [V/Fe] vs. [Fe/H] for the same stars in different analyses. The BACCHUS analyzed ARCES-stars are marked using blue squares, the \citet{2016ApJS..225...32B}-stars are marked using green diamonds, and the \citet{2015A&A...580A..24D}-stars are marked using brown crosses, the values from Gaia-ESO DR3 are marked using black open circles, and the APOGEE results are marked using black filled circles. In the bottom row panels, the values from \citet{2015A&A...577A...9B} are shown in the background using gray dots.\label{fig:v}}
\end{figure*}

All lines used in ASPCAP are sensitive to the adopted effective temperature, and as such the derived vanadium abundances are expected to be influenced by the effective temperature-trend with metallicity in DR13/DR14. When comparing to the references in the upper row of panels in Figure \ref{fig:v}, a positive trend with [Fe/H] indeed can be seen, especially in the Cannon analysis, which might be a product of the trend of effective temperature with metallicity.

Vanadium is produced in explosive oxygen and silicon burning, and on a cosmic scale, vanadium is produced by both SNeII and SNeIa in comparable amounts \citep{2003hic..book.....C}.

Regarding the [V/Fe] vs. [Fe/H] trends in the lower row of panels in Figure \ref{fig:v}, the only reference sample showing an obvious trend is the reference trend from \citet{2015A&A...577A...9B}. However, within the metallicity interval where comparison stars are available in any of the references ($-0.5\lesssim$[Fe/H]$\lesssim0.5$), all trends are fairly flat. The BACCHUS-analyzed ARCES-stars and \citet{2016ApJS..225...32B} possibly show a slight negative trend of [V/Fe] for [Fe/H]$\gtrsim0$, something that is not shown in the other samples. The trend found in \citet{2015A&A...577A...9B} is very alpha-like, and indeed they draw the conclusion that vanadium likely is mainly produced in SNeII, which is the main producer of the alpha elements.

\subsection{Chromium, Cr}
In ASPCAP, the chromium abundance is determined from eight regions covering Cr I lines.

\begin{figure*}
\epsscale{1.1}
\plotone{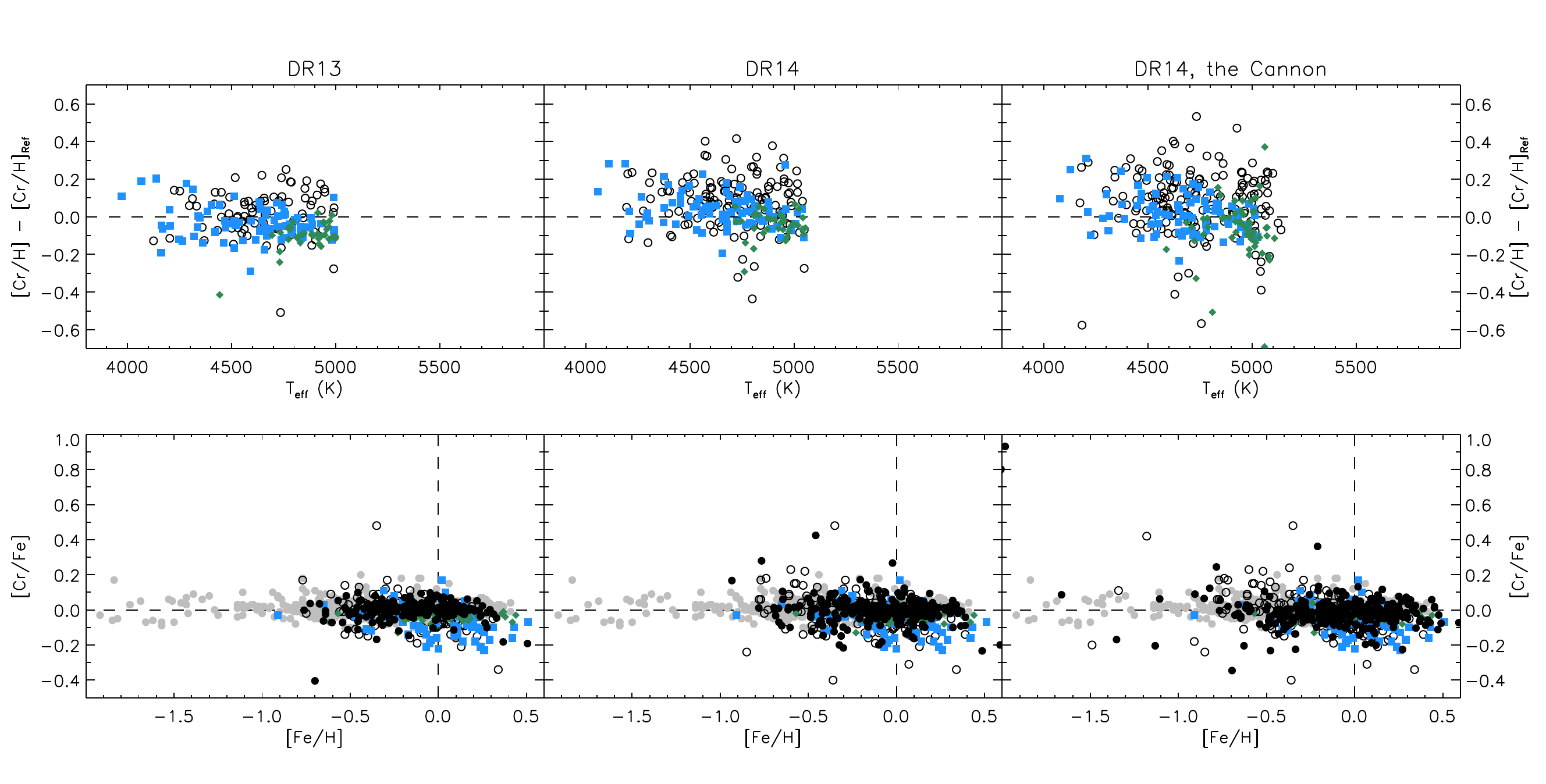}
\caption{The first row shows differences in chromium abundance for the same stars in different analyses, and the second row shows [Cr/Fe] vs. [Fe/H] for the same stars in different analyses. The BACCHUS analyzed ARCES-stars are marked using blue squares, and the \citet{2016ApJS..225...32B}-stars are marked using green diamonds the values from Gaia-ESO DR3 are marked using black open circles, and the APOGEE results are marked using black filled circles. In the bottom row panels, the values from \citet{2014A&A...562A..71B} are shown in the background using gray dots.\label{fig:cr}}
\end{figure*}

Compared with the references, the DR13 chromium abundances are 0.03 dex lower, with a spread of 0.11 dex. For DR14, the systematic shift is 0.04 dex and the spread is 0.12 dex, as illustrated in the first row of panels in Figure \ref{fig:cr}.

Chromium is produced in explosive silicon burning, and on a cosmic scale, SNeIa and SNeII contribute roughly equal parts of the total chromium budget \citep{2003hic..book.....C}. Its chemical evolution is expected to follow that of iron. Indeed, this is the case for all analyses in the [Cr/Fe] vs. [Fe/H] trends in the bottom row of panels in Figure \ref{fig:cr}.
Excluding some spurious outliers, the APOGEE trends seem tighter and more closely following that of \citet{2014A&A...562A..71B} than the trends of the BACCHUS analyzed ARCES-stars and the Gaia-ESO survey.

\subsection{Manganese, Mn}
In ASPCAP, the manganese abundance is determined from ten regions of the spectra covering (hfs) Mn I lines, but the three lines that are given the most weight in the manganese abundance determination are the (hfs) Mn I lines at 15159.2~\AA, 15217.8~\AA, and 15262.5~\AA.

\begin{figure*}
\epsscale{1.1}
\plotone{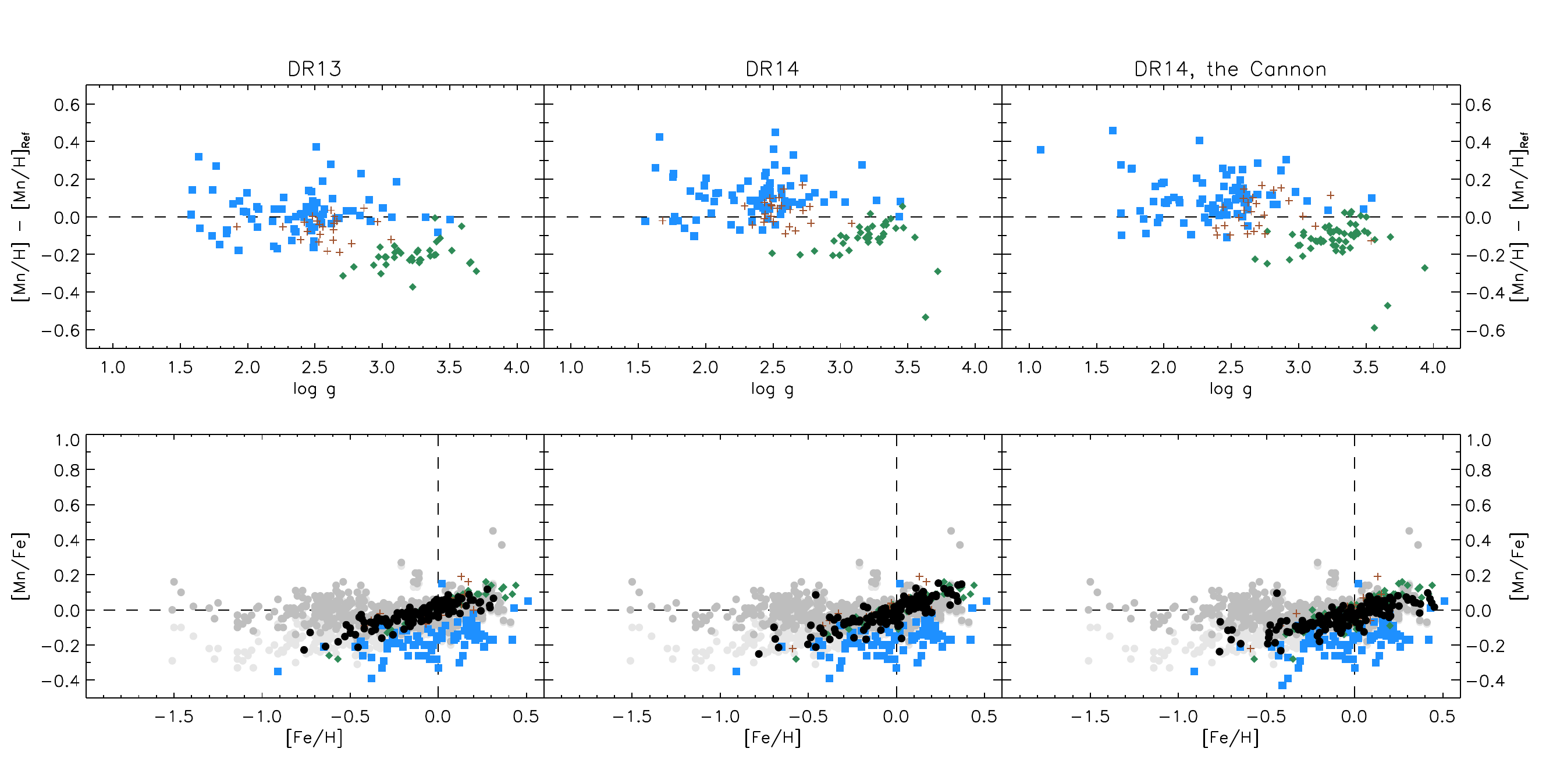}
\caption{The first row shows differences in manganese abundance for the same stars in different analyses, and the second row shows [Mn/Fe] vs. [Fe/H] for the same stars in different analyses. The BACCHUS analyzed ARCES-stars are marked using blue squares, the \citet{2016ApJS..225...32B}-stars are marked using green diamonds, the \citet{2015A&A...580A..24D}-stars are marked using brown crosses, and the APOGEE results are marked using black filled circles. In the bottom row panels, the LTE-values from \citet{2015A&A...577A...9B} are shown in the background using light gray dots, and their NLTE-corrected values are shown using dark gray dots.\label{fig:mn}}
\end{figure*}

Compared with the references, the DR13 manganese abundances are 0.03 dex lower than the references with a spread of 0.14 dex, while for DR14, the systematic shift is 0.05 dex and the spread is 0.14 dex, see the first row of panels in Figure \ref{fig:mn}.

Manganese is produced by explosive silicon burning and in the alpha-rich freezeout in SNeII, and on a cosmic scale, both SNeIa and SNeII contribute significant parts of the manganese budget \citep{2003hic..book.....C}.

In the bottom row of Figure \ref{fig:mn}, the light gray dots mark the LTE-results of \citet{2015A&A...577A...9B}, and the darker gray dots mark their NLTE-corrected abundances (from Bergemann, priv. comm.). The APOGEE-trends all show very tight trends closely following the LTE-trend of \citet{2015A&A...577A...9B}. This suggests that the manganese abundances of APOGEE might be very precise, but may need to be corrected for NLTE effects. We note that the [Mn/Fe] vs. [Fe/H] trend from the BACCHUS analyzed ARCES-stars is much lower and more scattered than the APOGEE-trend. 

\subsection{Iron, Fe}
In ASPCAP, the iron abundance is determined from numerous windows in the spectra covering Fe I lines.

In this comparison work, [Fe/H] is considered one of the stellar parameters and hence discussed in Section \ref{sec:params}.

\subsection{Cobalt, Co}
In ASPCAP, the cobalt abundance is determined from one (hfs) Co I line at 16757.6~\AA, and three more regions with blended and weak lines (in GK-giants). The 16757.6~\AA-line is given the highest weight in the ASPCAP windows, and in practice drives the cobalt abundance determination.

Cobalt is produced by explosive silicon burning, in the alpha-rich freezeout in SNeII, and by the s-process. On a cosmic scale, both SNeIa and SNeII contribute significant parts of the cobalt budget \citep{2003hic..book.....C}.

\begin{figure*}
\epsscale{1.1}
\plotone{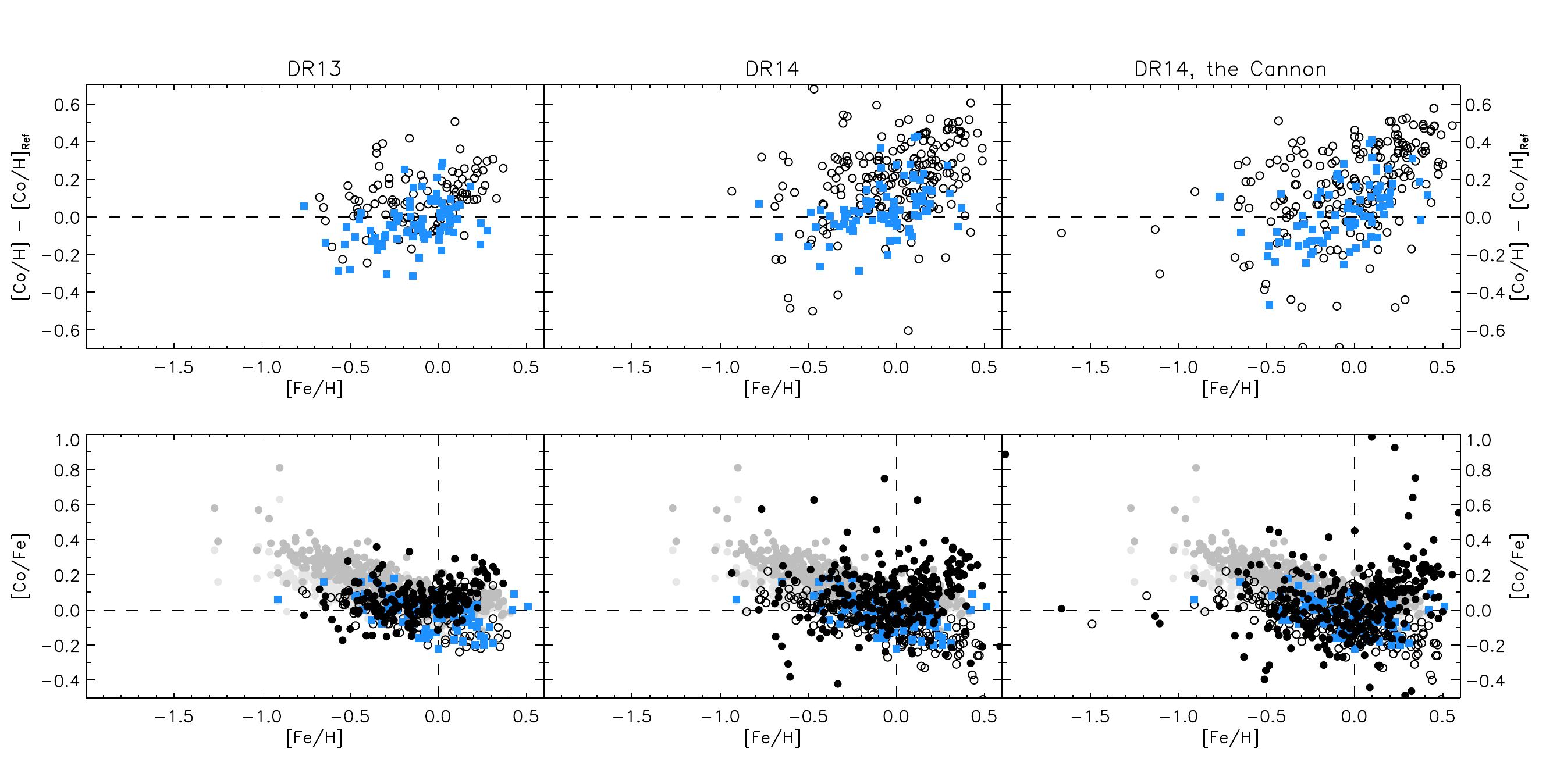}
\caption{The first row shows differences in cobalt abundance for the same stars in different analyses, and the second row shows [Co/Fe] vs. [Fe/H] for the same stars in different analyses. The BACCHUS analyzed ARCES-stars are marked using blue squares, the values from Gaia-ESO DR3 are marked using black open circles, and the APOGEE results are marked using black filled circles. In the bottom row panels, the LTE-values from \citet{2015A&A...577A...9B} are shown in the background using light gray dots, and their NLTE-corrected values are shown using dark gray dots.\label{fig:co}}
\end{figure*}

From the top row of panels in Figure \ref{fig:co}, one can make out a trend of cobalt abundance difference between the APOGEE and reference values with metallicity, which is also seen as the upturn for the most metal-rich part of the [Co/Fe] vs. [Fe/H] trends in the bottom row of panels. From the bottom row of plots, one can also note that the optical trends from the BACCHUS analyzed ARCES-stars and the Gaia-ESO survey do not follow that of \citet{2015A&A...577A...9B}.

As noted in \citet{paperi}, the ASPCAP-derived cobalt abundances show strong effective temperature trends within clusters, and should be used with caution. 

\subsection{Nickel, Ni}
In ASPCAP, the nickel abundance is determined from 30 regions of the spectra covering Ni I lines, several of which seem unblended and of suitable strength, while some are weak and blended (in GK-giants).

\begin{figure*}
\epsscale{1.1}
\plotone{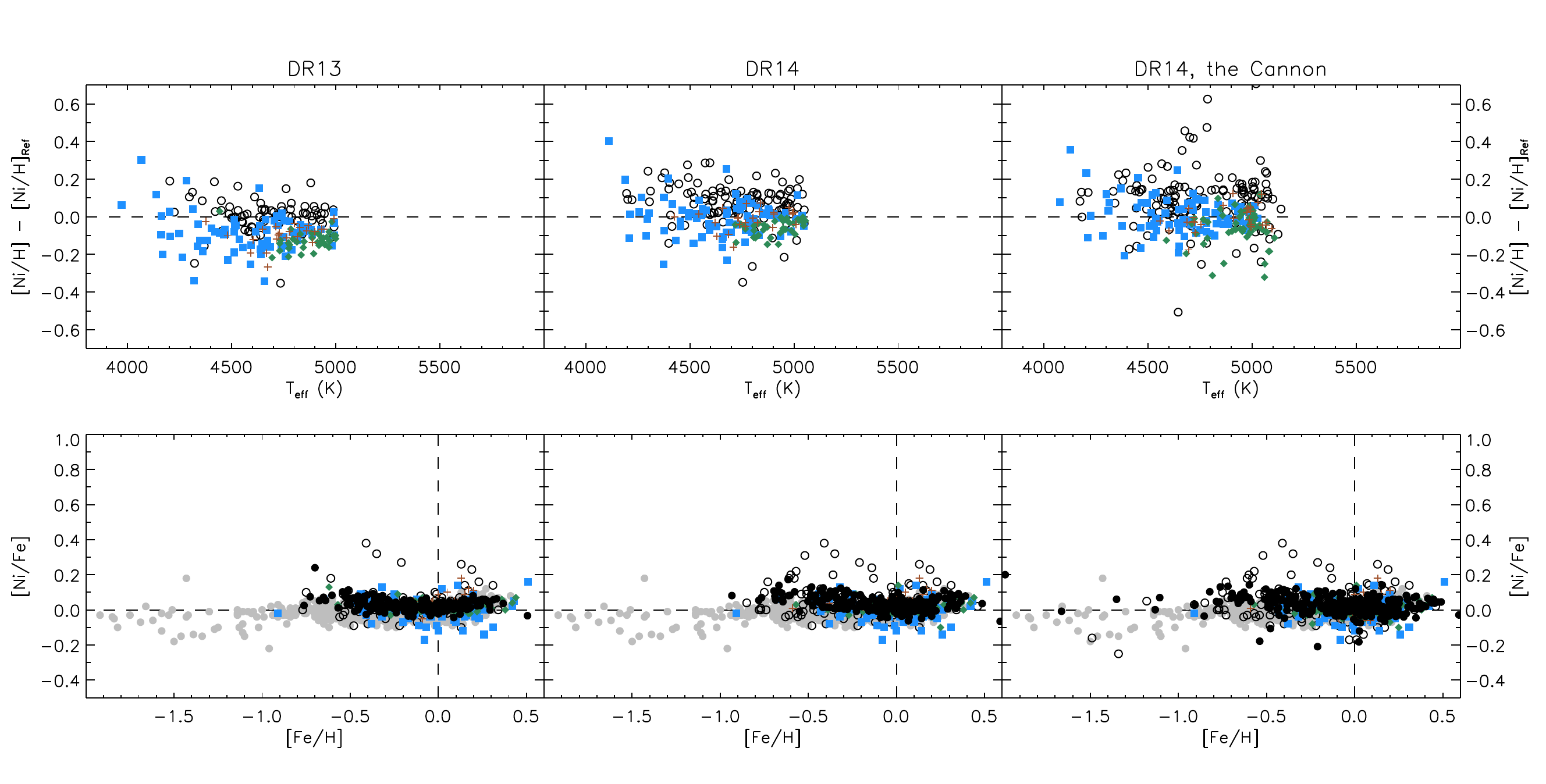}
\caption{The first row shows differences in nickel abundance for the same stars in different analyses, and the second row shows [Ni/Fe] vs. [Fe/H] for the same stars in different analyses. The BACCHUS analyzed ARCES-stars are marked using blue squares, the \citet{2016ApJS..225...32B}-stars are marked using green diamonds, and the \citet{2015A&A...580A..24D}-stars are marked using brown crosses, the values from Gaia-ESO DR3 are marked using black open circles, and the APOGEE results are marked using black filled circles. In the bottom row panels, the values from \citet{2014A&A...562A..71B} are shown in the background using gray dots.\label{fig:ni}}
\end{figure*}

The DR13 nickel abundances are 0.07 dex lower than the references with a spread of 0.10 dex. For DR14, the systematic shift is +0.02 dex and the spread is 0.10 dex; see the first row of panels in Figure \ref{fig:ni}.

Nickel is produced by explosive silicon burning, in the alpha-rich freezeout in SNeII, and by the weak s-process. On a cosmic scale, both SNeIa and SNeII contribute significant parts of the nickel budget \citep{2003hic..book.....C}.

The chemical evolution of nickel is expected to follow that of iron. Indeed, this is the case for all analyses in the [Ni/Fe] vs. [Fe/H] trends in the bottom row of panels in Figure \ref{fig:ni}. All APOGEE analyses show very tight trends and nickel is the most accurate iron-peak element when compared to the references.

\subsection{Neutron capture elements}\label{sec:nc}
The line list used for analysis in DR13 and DR14 did not include many transitions from neutron capture elements. All lines from such elements are weak and/or blended, and the current analysis methodology is likely to have significant challenges with these. As a result, DR13/DR14 abundances for these elements are not likely to be valid, and no calibrated abundances are given in DR14 (DR13 erroneously populated the calibrated arrays for Cu, Ge, Rb, and Y(Yb)); hence no comparisons are made here. Work is ongoing to improve the analysis. We describe the status for several elements below. Note that there are still unidentified lines in APOGEE spectra that may yield more possibilities.

\begin{center}
{\it Copper, Cu}
\end{center}

In ASPCAP, a copper abundance determination is attempted from the heavily blended hfs Cu I line at 16005.8~\AA~and a very weak (in GK-giants) hfs Cu I line at 16639.0~\AA.

\citet{2016A&A...594A..43H} use the (hfs) Cu I lines at 16005.8~\AA~and 16006.6~\AA, but only quote upper limits on the copper abundance. \citet{2013ApJ...765...16S} use only the line at 16005.8~\AA.

Further investigations will be made of these lines and their possible utility in APOGEE spectra before the next data release.

\begin{center}
{\it Germanium, Ge}
\end{center}

A germanium abundance determination is attempted in ASPCAP from a very weak (in GK-giants) Ge I line at 16759.8~\AA~that is heavily blended with a Fe I line at the resolution of APOGEE. As noted in \citet{paperi}, the ASPCAP-derived germanium abundances show strong effective temperature trends within clusters, and should be used with caution. Neither \citet{2013ApJ...765...16S} nor \citet{2016A&A...594A..43H} attempt to determine the germanium abundance.

Further investigations will be made of this line and its possible utility in APOGEE spectra before the next data release.

\begin{center}
{\it Rubidium, Rb}
\end{center}

In ASPCAP, a rubidium abundance determination is attempted from one weak (in GK-giants), heavily blended (with Fe I) Rb I line at 15289.5~\AA.

\citet{2016A&A...594A..43H} use the same line, but only quote upper limits on the rubidium abundance, while \citet{2013ApJ...765...16S} do not measure rubidium at all.

Recent investigations of this line by the ASPCAP team have shown that it is too weak to be useful in the vast majority of the APOGEE-spectra, and in fact the rubidium abundances have been removed in DR14.

\begin{center}
{\it Cerium, Ce}
\end{center}

In coming versions of ASPCAP, cerium abundances will be determined from nine Ce II lines, see \citet{2017ApJ...844..145C} for details. The lines are relatively unblended and of suitable strengths (in GK-giants), and tests made in \citet{2017ApJ...844..145C} suggest that cerium will be measurable in the bulk of APOGEE spectra through these lines. Cerium abundances are planned to be included in the next data release.

\begin{center}
{\it Neodymium, Nd}
\end{center}

In ASPCAP, the neodymium abundances are presently derived using one very weak (in GK-giants), blended Nd II line at 16053.6~\AA~covering only two data points in the reduced spectrum. In upcoming versions, however, the ten Nd II lines described in \citet{2016ApJ...833...81H} will be added to the line list. Based on tests made in \citet{2016ApJ...833...81H}, these new lines will allow neodymium abundances to be reliably determined in about 18\% of the APOGEE red giants.

\begin{center}
{\it Ytterbium, Yb}
\end{center}

In ASPCAP, the quoted yttrium (Y) abundance in fact is the ytterbium (Yb) abundance, and it is derived from a weak (in GK-giants), CO-blended Yb II line 16498.4~\AA. \citet{2016A&A...594A..43H} use the same line to derive upper limits on the ytterbium abundance.

Further investigations will be made of this line and its possible use in APOGEE spectra until the next data release.

\section{Conclusions}\label{sec:conclusions}
In this paper, we have compared the stellar parameters and abundances of independent optical studies to those presented by the APOGEE DR13 and DR14 (including the Cannon) on a star-by-star basis. We choose to only make the comparisons using stars with calibrated values supplied for all three `classic' spectroscopic stellar parameters in the APOGEE data releases, which leads us to restrict the comparison to subgiant and giant stars, and exclude dwarf stars. Since the giant stars are the main targets of the APOGEE survey, this approach leads to the best evaluation of the general performance of the analysis pipeline for the bulk of the surveyed stars.

For the stellar parameters, $\log g$ and [Fe/H] match the optical works well, as is shown in Table \ref{tab:params}. However, we have found that the effective temperatures in DR13, DR14, and in the DR14-analysis using the Cannon show trends with the metallicity of the star. The calibrated effective temperatures given in DR14 are better than those of DR13, and agree very well with the reference values for $-1.0\lesssim$[Fe/H]$\lesssim-0.5$. For higher metallicities, the DR14 effective temperatures are too high by the order of 100 K, and the behavior is unclear for [Fe/H]$<-1.0$ since there are very few reference values for these low metallicities.

For most of the elements -- C, Na, Mg, Al, Si, S, Ca, Cr, Mn, Ni -- the DR14 ASPCAP analysis have systematic differences to the comparison samples of less than 0.05 dex (median), and random differences of less than 0.15 dex (standard deviation). Compared to the comparison samples, magnesium is the alpha element for which we find best consistency. The ASPCAP [Mg/Fe] vs. [Fe/H] trend shows a clear thin/thick disk separation, is tight, and very similar to the reference works. When it comes to iron-peak elements, nickel is the most accurate element in APOGEE compared to the references (besides iron itself, that is).

When it comes to elements formed by the r- and s-processes, work will be done to evaluate the possibilities to determine copper, germanium, and/or ytterbium abundances in coming data releases, and work is already ongoing on determining cerium and neodymium abundances for a majority of the observed stars, and they are planned to be released in the next data release.

The abundances of some elements -- N, O, K, Ti I, V, Co -- show strong correlations with some determined stellar parameter when comparing to the reference studies. Some of these trends might be due to the trend of determined effective temperature with metallicity in ASPCAP and the fact that these uncalibrated stellar parameters are used when subsequently determining the stellar abundances. Our tests have shown that this is the case for Ti I, but it is still uncertain whether the same can be said for the other elements. Regarding oxygen, for example, our tests have shown that the situation gets worse if this change in methodology is invoked. This will be investigated further for the next data release, but for now we note that even if the accuracy of the ASPCAP-derived oxygen abundance might be in need of improvement, its precision is very high for stars of similar types \citep{2016A&A...590A..74B}.

The best way to remove the impact of the effective temperature trend would be to identify and remove the source of the trend in the ASPCAP analysis; this might for example be accomplished by updating line lists and/or using specific windows in the spectra for the determination of the effective temperature (and/or the other stellar parameters). To better understand the origin and complete impact of the trend, more overlap between independently-analyzed metal poor stars is needed for the next data release. This could be accomplished by either targeting metal-poor giants from the work of \citet{2011ApJ...737....9R} with the APOGEE instrument, or by observing and analyzing already targeted metal-poor APOGEE-stars using another spectrometer. Preferably both approaches should be used to make certain that any possible trend seen is not due to any possible systematics in the analysis of \citet{2011ApJ...737....9R}. 

Regarding the growing fraction of FGK-dwarf stars from DR13 to DR14 (from 26\% to 34\%), it would also be desirable to have more such stars in common between APOGEE DRs and independent analyses, to enable a evaluation of the performance of ASPCAP and possibly calibrate the results in this region of the HR diagram. There is a multitude of such studies available \citep[][etc.]{2003MNRAS.340..304R,2006MNRAS.367.1329R,2014A&A...562A..71B}, some for which the overlap with APOGEE is already quite large \citep{2016ApJS..225...32B,2017AJ....154..107P}. However, care must be taken to target a wide range of metallicities of these overlapping stars during coming observations with APOGEE so that any possible trend with metallicity could be traced.

\acknowledgments
H. J\"onsson acknowledges support from the Birgit and Hellmuth Hertz’ Foundation (via the Royal Physiographic Society of Lund), the Crafoord Foundation, and Stiftelsen Olle Engkvist Byggm\"astare. C. Allende Prieto is thankful to the Spanish Ministry of Economy and Competitiveness (MINECO) for support through grant AYA2017-86389-P. D. K. Feuillet acknowledges funds from the Alexander von Humboldt Foundation in the framework of the Sofja Kovalevskaja Award endowed by the Federal Ministry of Education and Research. K. Hawkins is partially funded by the Simons Foundation Society of Fellows and the Flatiron Institute Center for Computational Astrophysics in New York City. S. M{\'e}sz{\'a}ros has been supported by the Premium Postdoctoral Research Program of the Hungarian Academy of Sciences, and by the Hungarian NKFI Grants K-119517 of the Hungarian National Research, Development and Innovation Office. D. A. Garc\'ia-Hern\'andez and O. Zamora acknowledge support provided by the Spanish Ministry of Economy and Competitiveness (MINECO) under grant AYA-2017-88254-P. J. G. Fern\'andez-Trincado is supported by FONDECYT No. 3180210.

Funding for the Sloan Digital Sky Survey IV has been provided by the Alfred P. Sloan Foundation, the U.S. Department of Energy Office of Science, and the Participating Institutions. SDSS acknowledges support and resources from the Center for High-Performance Computing at the University of Utah. The SDSS web site is www.sdss.org.

SDSS is managed by the Astrophysical Research Consortium for the Participating Institutions of the SDSS Collaboration including the Brazilian Participation Group, the Carnegie Institution for Science, Carnegie Mellon University, the Chilean Participation Group, the French Participation Group, Harvard-Smithsonian Center for Astrophysics, Instituto de Astrof\'isica de Canarias, The Johns Hopkins University, Kavli Institute for the Physics and Mathematics of the Universe (IPMU) / University of Tokyo, Lawrence Berkeley National Laboratory, Leibniz Institut f\"ur Astrophysik Potsdam (AIP), Max-Planck-Institut f\"ur Astronomie (MPIA Heidelberg), Max-Planck-Institut f\"ur Astrophysik (MPA Garching), Max-Planck-Institut f\"ur Extraterrestrische Physik (MPE), National Astronomical Observatories of China, New Mexico State University, New York University, University of Notre Dame, Observat\'orio Nacional / MCTI, The Ohio State University, Pennsylvania State University, Shanghai Astronomical Observatory, United Kingdom Participation Group, Universidad Nacional Aut\'onoma de M\'exico, University of Arizona, University of Colorado Boulder, University of Oxford, University of Portsmouth, University of Utah, University of Virginia, University of Washington, University of Wisconsin, Vanderbilt University, and Yale University.

This publication made use of the SIMBAD database, operated at CDS, Strasbourg, France, and NASA's Astrophysics Data System.

\facilities{ARC (ARCES), Sloan (APOGEE)}

\software{
ASPCAP \citep{2016AJ....151..144G},
ATLAS9 (Kurucz 1979, Castelli \& Kurucz 2003), 
Brussels Automatic Code for Characterizing High AccUracy Spectra  (BACCHUS) \citep{2016ascl.soft05004M}, 
FERRE  \citep{2006ApJ...636..804A}, 
MARCS \citep{2008A&A...486..951G}, 
Turbospectrum \citep{1998A&A...330.1109A,2012ascl.soft05004P}
}


\end{document}